\documentclass[journal]{IEEEtran}

\IEEEoverridecommandlockouts

\newcommand{\unit}[1]{\ensuremath{\, \mathrm{#1}}}

\newcommand{\pct}{\%}
\newcommand{\hlchange}[1]{\textcolor{black}{#1}}
\newcommand{\hlchangeB}[1]{\textcolor{black}{#1}}

\ifCLASSINFOpdf
  \usepackage[pdftex]{graphicx}
\fi

\usepackage[cmex10]{amsmath}
\usepackage[tight,footnotesize]{subfigure}

\usepackage{url}

\usepackage{epsfig}
\usepackage{cite} 
\usepackage{threeparttable}
\usepackage{pgfplots}
\usepackage{enumerate}
\usetikzlibrary{pgfplots.groupplots}
\pgfplotsset{grid style={dashed}}
\usepgfplotslibrary{smithchart}
\usetikzlibrary{backgrounds}
\usepackage{helvet}
\usepackage[eulergreek]{sansmath}
\pgfplotsset{
  tick label style = {font=\sansmath\sffamily},
  every axis label = {font=\sansmath\sffamily},
  legend style = {font=\sansmath\sffamily},
  label style = {font=\sansmath\sffamily}
}

\begin{document}
\bstctlcite{IEEEexample:BSTcontrol}
%
\title{Energy-Efficient Low-Power Circuit Techniques for Wireless Energy and Data Transfer in IoT Sensor Nodes}

\author{Gustavo C. Martins, ̃\IEEEmembership{Student Member, ̃IEEE,} Alessandro Urso, \IEEEmembership{Student Member, ̃IEEE,} Andr\'e Mansano, \IEEEmembership{Member, ̃IEEE,} Yao Liu and Wouter A. Serdijn, ̃\IEEEmembership{Fellow, ̃IEEE}%
\thanks{G. C. Martins, A. Urso, and W. A. Serdijn are with Delft University of Technology, Section Bioelectronics, Mekelweg 4, 2628 CD  Delft, the Netherlands (email: g.c.martins@ieee.org, a.urso@tudelft.nl, serdijn@ieee.org). A. Mansano was with Delft University of Technology, Section Bioelectronics and is now with Nowi Energy, Delft, the Netherlands (email: amansano@gmail.com). Y. Liu was with Delft Univerty of Technology, Section Bioelectronics and is now with IMEC, Leuven, Belgium (email: yaoliuhust@gmail.com).}
\thanks{Part of this work was supported by CNPq and CAPES Foundations, Brazil, and the China Scholarship Council.}
\thanks{Copyright (c) 2017 IEEE. Personal use of this material is permitted.
However, permission to use this material for any other purposes must be
obtained from the IEEE by sending an email to pubs-permissions@ieee.org}
}


\maketitle

\begin{abstract}
In this paper, we present techniques and examples to reduce power consumption and increase energy efficiency of autonomous Wireless Sensor Nodes (WSNs) for the Internet of Things.
\hlchange{We focus on the RF Energy Harvester (RFEH), the data receiver and the transmitter, all of which have a large impact on the device cost, lifetime and functionality.
Co-design of the antenna and the electronics is explored to boost the power conversion efficiency of the RF-DC converter.
\hlchangeB{As a proof of principle, a} charge pump rectifier is designed, and its measurement results are presented.
To boost the rectifier output voltage, a DC-DC converter that employs maximum power point tracking has been designed. \hlchangeB{A prototype circuit is also presented that can accommodate an} input power level range of $1 \unit{\mu W}$ to $ 1 \unit{mW}$ and offers peak efficiencies  of $76.3\%$  and $82\%$  at $1 \unit{\mu W}$ and $1 \unit{mW}$, respectively.
The co-design principle is also used at the receiver side where the antenna-electronics interface is optimized. 
It is shown how this technique allows improving the noise figure of the Low Noise Amplifier (LNA) without sacrificing power consumption. 
\hlchangeB{As a low power alternative to narrow-band wireless transmission, sub-GHz ultra-wideband is proposed. As a proof of principle,} the design of a novel low-power sub-GHz Ultra-Wide-Bandwidth (UWB) transmitter which consumes only $0.28 \unit{mW}$ is presented. Its working principle is  verified by means of circuit simulations and measurements. 
The low power nature of the transmitter \hlchangeB{and receiver principles}, combined with the power efficient RF-DC converter \hlchangeB{paves the way} towards continuous operation of a WSN.
}
\end{abstract}


\section{Introduction}
\label{Introduction}
\IEEEPARstart{W}{ith}
\hlchange{
 the development of the Internet of Things (IoT), the need for low cost Wireless Sensor Nodes (WSNs) is becoming larger and larger.
One expensive component, still used in the majority of WSNs, is the battery.
Additionally, the disposal of batteries is an expensive and environment unfriendly process, and sometimes the cost of batteries is even higher than the cost of the electronics involved.}
Furthermore, WSNs that require an eventual replacement of batteries are not suitable to be used in areas where human access is very limited.
For these reasons there is the need for remotely-powered battery-less WSNs. 

\begin{figure}
\includegraphics[width=0.45\textwidth]{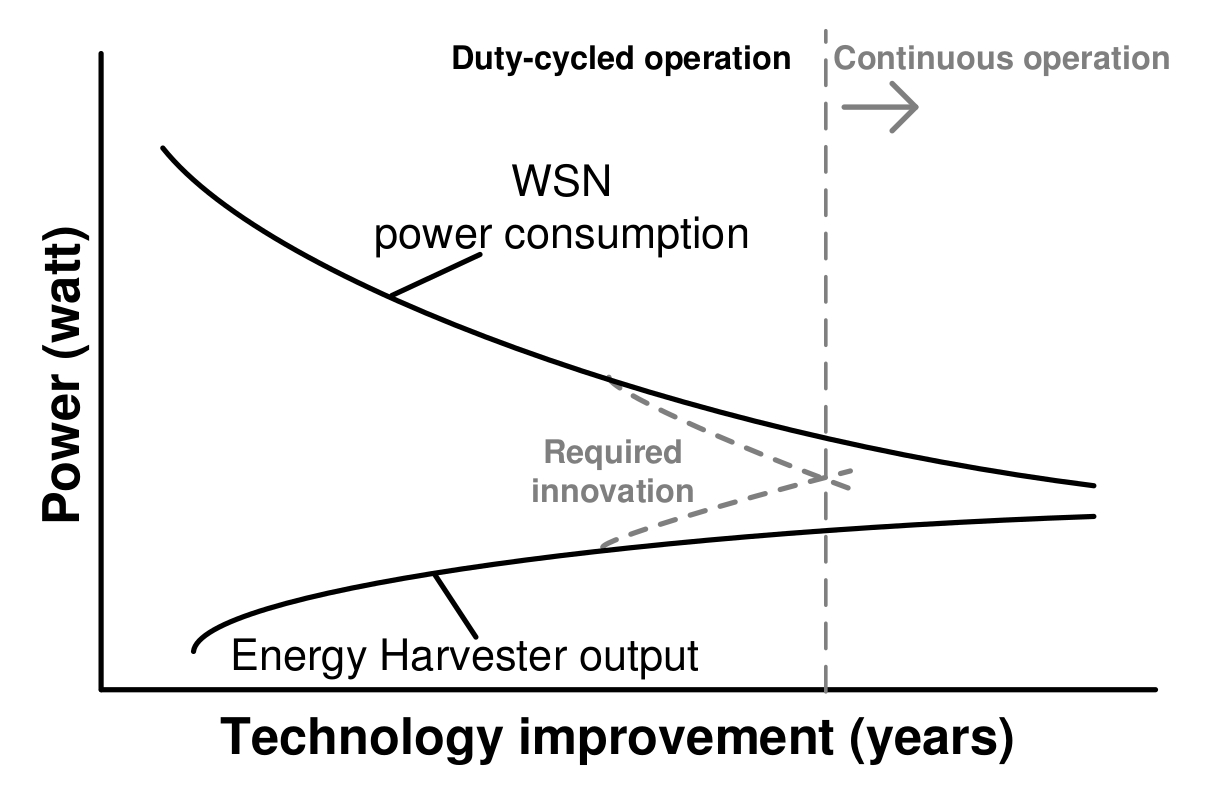}
  \caption{\hlchange{Forecast of the power balance evolution of a typical WSN \cite{Mark_thesis}.}}
  \label{power_scenario}
\end{figure}

\hlchange{
With technology improvement, the power consumption of WSNs tends to reduce whereas the efficiency of Energy Harvesters (EHs) tends to increase, as depicted in Fig. \ref{power_scenario}.
However, the gap between the power required by the electronics and the energy stored cannot yet be closed for most applications. Hence, continuous operation is not possible and the device needs to be duty-cycled.}
To further illustrate this, Tables \ref{tab:txpwr} and \ref{tab:ehpwr} present the typical power necessary for data transmission (the task that usually requires most of the power in a WSN) and the input available power of state-of-the-art energy harvesters, respectively.
Further improvements in energy harvesting and data transfer must be achieved to close this gap \hlchange{ and facilitate continuous operation.}

\begin{figure}
\centering
\includegraphics[width=0.4\textwidth]{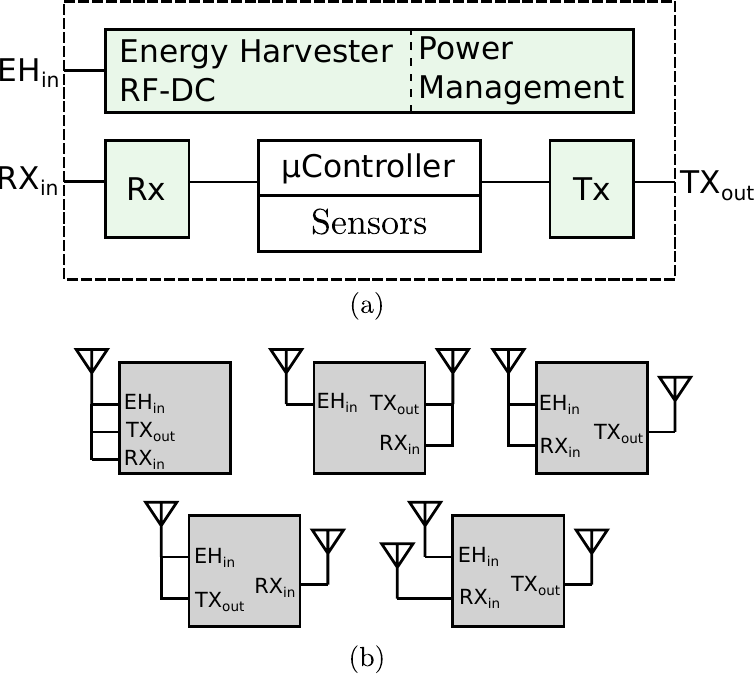}
  \caption{(a) Detailed block diagram of an autonomous wireless sensor node. (b) 5 possible configurations based on the number of available antennas.}
  \label{architecture}
\end{figure}

\begin{table}
\centering
\caption{Typical data transmission power consumption.}
\label{tab:txpwr}
\begin{tabular}{c|c}
Work & Power consumption\\
\hline
\cite{Martynenko2012} & $123 \unit {mW}$ (peak value) \\
\cite{Odeh2014} & $1.3 \unit{mW}$ \\
\cite{Wong2013} & $1.7 \unit{mW}$  
\end{tabular}
\end{table}

\begin{table}
\centering
\textcolor{black}{
\caption{State-of-the-art energy harvesters input power \hlchangeB{range and peak efficiency.}}
\label{tab:ehpwr}
\begin{tabular}{c|c|c}
Work & Input power range& Peak eff.\\
\hline
\cite{Mark_JSSC} & $\sim 160 \unit{nW}$-$19 \unit{\mu W}$ & $40\%$ \\
\cite{Hsieh2015} & $\sim 19 \unit{\mu W}$-$126 \unit{\mu W}$ & $44.1\%$ \\
\cite{Papotto2011} & $4 \unit{\mu W}$-$13 \unit{\mu W}$ & $11.5\%$ 
\end{tabular}
}
\end{table}

In Fig. \ref{architecture}, a general architecture of a typical autonomous WSN is depicted.
\hlchange{In this work, we focus on wireless energy harvesting as the method to remotely power the device.}
When powered, the microcontroller reads and processes data from sensors, which can subsequently be sent to a transmitting antenna.
\hlchange{Data can also be received and further processed by the microcontroller unit.
In order to wirelessly transmit and receive data and receive power, an antenna or multiple antennas are needed.
Since there are three RF input/output ports, the WSN can have up to three antennas.}
As depicted in Fig. \ref{architecture}(b), based on how the available antennas are connected, 5 different scenarios can be considered:
\begin{enumerate}[a)]
\item A WSN having one antenna. The antenna is shared between the transmitter, the receiver and the EH. In this scenario, these three blocks all have to work at the same frequency.
\item A WSN having two antennas. One antenna is used by the EH, while the other is shared between the transmitter and the receiver.
\item A WSN having two antennas. One antenna is used by the transmitter, while the other is shared between the receiver and the EH.
\item A WSN having two antennas. One antenna is used by the receiver, while the other is shared between the transmitter and the EH.
\item A WSN having three antennas: one antenna for the receiver, one for the transmitter and one for the EH.
\end{enumerate}
Scenario e) has the most degrees of freedom. This means that the most suitable frequency for each block can be used. 
For instance, the energy can be received at lower frequencies to achieve higher efficiency, while the data can be transferred at higher appropriate frequencies to achieve higher data rates.
\hlchangeB{However, despite having the most degrees of freedom, Scenario e) is also the option that presents the highest cost and largest physical dimensions, due to the additional antennas.}

The purpose of this paper is, irrespective of the number of available antennas, to present and discuss circuit techniques that allow to bring the two curves of Fig. \ref{power_scenario} as close as possible or even cross each other.

This paper is organized as follows. 
\hlchange{Each highlighted block of Fig. \ref{architecture}(a) will be discussed in its own section, from Sections \ref{sec:pwr} to \ref{sec:datatx}.
Each section is divided in three subsections: \textit{Circuit Description}, \textit{Implementation and Results} and \textit{Discussion}.
In the \textit{Circuit Description} subsection, the idea and the design principles are illustrated.
As a proof of concept, in the \textit{Implementation and Results} subsection, a circuit is designed and its working principle is verified by means of circuit simulations or measurements.
In the \textit{Discussion} subsection, some final comments are given. 
Section \ref{sec:pwr} discusses the RF to DC conversion and the power management unit.
A topology of a charge pump rectifier and its design are presented.
To boost the rectifier output voltage and charge a storage capacitor, it employs a DC-DC converter with Maximum Power Point Tracking (MPPT).
This allows to change the input resistance of the DC-DC converter in order to maximize the energy conversion efficiency of the power conversion chain.
In Section \ref{sec:datarx}, the concept of co-designing the antenna and the electronics for data reception is presented.
Unlike the co-design of the antenna and the rectifier in a \hlchangeB{Radio-Frequency Energy Harvester (RFEH)}, we optimize the interface between the antenna and the LNA in this section. 
Section \ref{sec:datatx} focuses on data transmission.
The design of a novel low-voltage low-power sub-GHz Ultra-Wide-Bandwidth (UWB) transmitter is presented. 
Its mathematical analysis is provided and its correct operation is validated by means of circuit simulations and measurements.
Finally, in Section \ref{sec:conc}, a summary of the paper and conclusions are given.}

\section{RF Energy Harvesting and Power Management}
\label{sec:pwr}
RF energy harvesting can reduce costs of WSNs, enabling many applications in the IoT domain.
\hlchange{However, its output power may be not high enough to power most applications.}
To increase the output power of such a harvester, the efficiency of the entire power conversion chain must be optimized.
\hlchange{Here we explore the optimization of the blocks that compose the power conversion chain, viz. the antenna-rectifier interface, rectifier design and rectifier-load interface, keeping in mind that all stages contribute to the overall efficiency.}

\hlchange{
\subsection{Circuit Description}}
To maximize the power transfer from the antenna to the rectifier, their impedances must be matched.
If an IC is directly connected to an off-chip antenna and the length
between them is electrically short, the antenna and the circuitry
can be directly matched without any intermediate stage(s). An optimum
choice of antenna impedance $Z_A$ and load impedance
$Z_L$ allows us to increase the voltage or current at
the antenna load for the same available power at the antenna. As addressed
in \cite{Mark_TCAS2},\emph{ }one needs to design the electronic circuit
for the largest quality factor of $Z_L$ possible and
subsequently co-design the antenna impedance for conjugate matching\emph{.}
This conclusion is a key point that needs to be considered during
a co-design procedure. If $Z_A$ is conjugately matched
with $Z_L$, and the antenna resistance $R_A$
is much smaller than the antenna reactance $X_A$, the
load voltage can be approximated as \hlchange{\cite{Mark_TCAS2}}:
\begin{equation}
\left.V_{L}\right|_{R_{A}\ll X_{A}}\approx\sqrt{2P_{av}}\frac{X_{A}}{\sqrt{R_{A}}},\label{eq:VoutBoost}
\end{equation}
which suggests that the output voltage is passively boosted by the
presence of the antenna reactance, which forms an LC resonator with
the load. Significant improvement of the rectifier input voltage for large values of Q can be achieved
at the expense of bandwidth. This property is exploited in \cite{Mark_JSSC},
where the input voltage of the RF energy harvester is effectively
increased using a high-Q loop antenna. This voltage boost improves the rectifier sensitivity,
meaning that a wireless sensor node with an RF energy harvester can
be operated at a larger distance from the RF energy source. 

In summary, the two conditions that need to be met for an optimum co-design are
as follows. The first condition is to conjugate match the antenna-electronics
interface as this maximizes both the voltage and current at the load.
The second condition is related to the fact that, since the available input power and the antenna load are fixed, the voltage at the load cannot be increased to a higher value only by means of proper antenna design. Therefore, one should determine at which impedance level conjugate matching should occur in order to further increase the load voltage or current. 

\begin{figure} 
\centerline{\psfig{figure=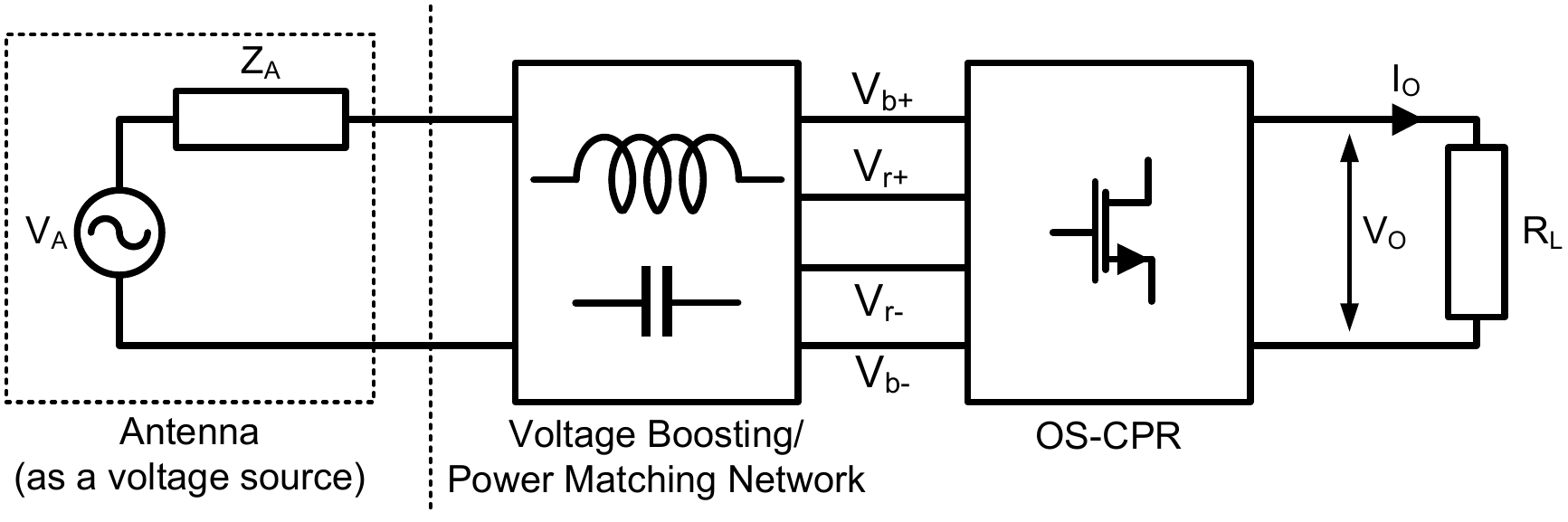,width=0.97\columnwidth}}
\caption{\small Block diagram of the RF energy harvester~\cite{AM}.}
\label{RFEHBD}
\end{figure}

\begin{figure}
\begin{center}
\subfigure[]{
\hspace{-1mm}\centerline{\psfig{figure=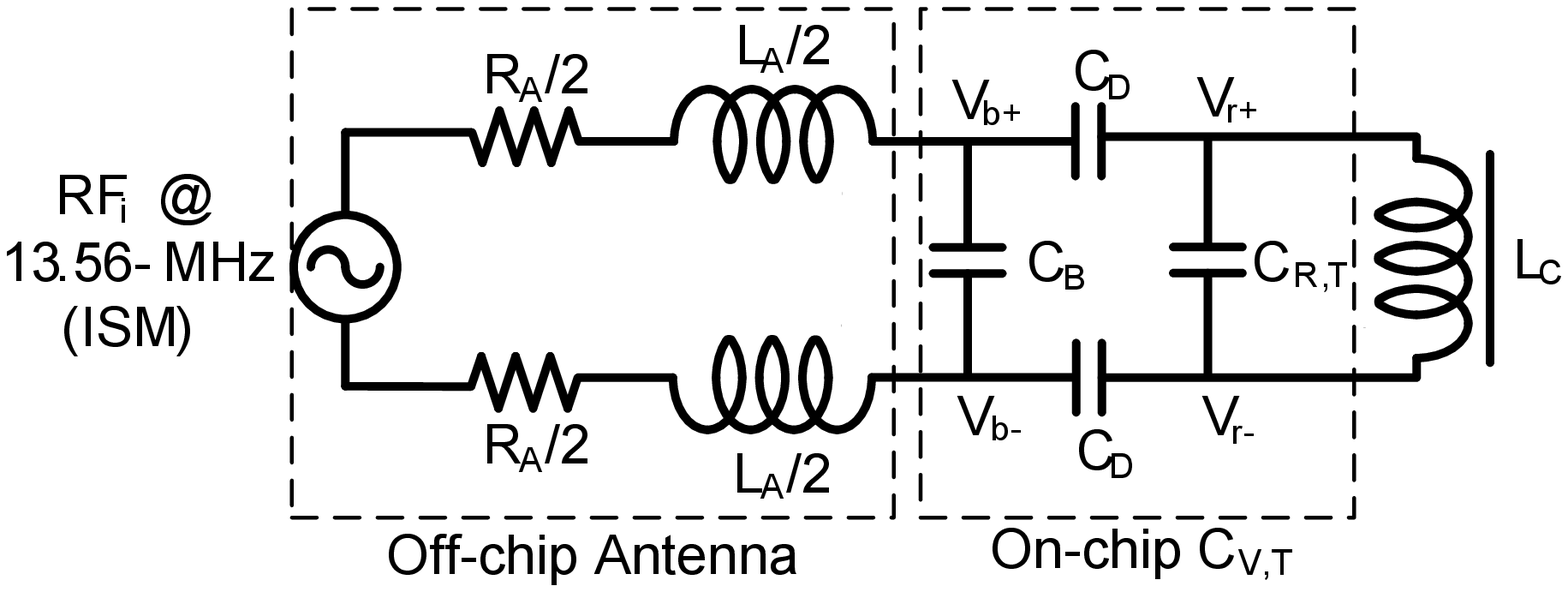,width=0.99\columnwidth}}
\label{VBN}
}
\subfigure[]{
\centerline{\psfig{figure=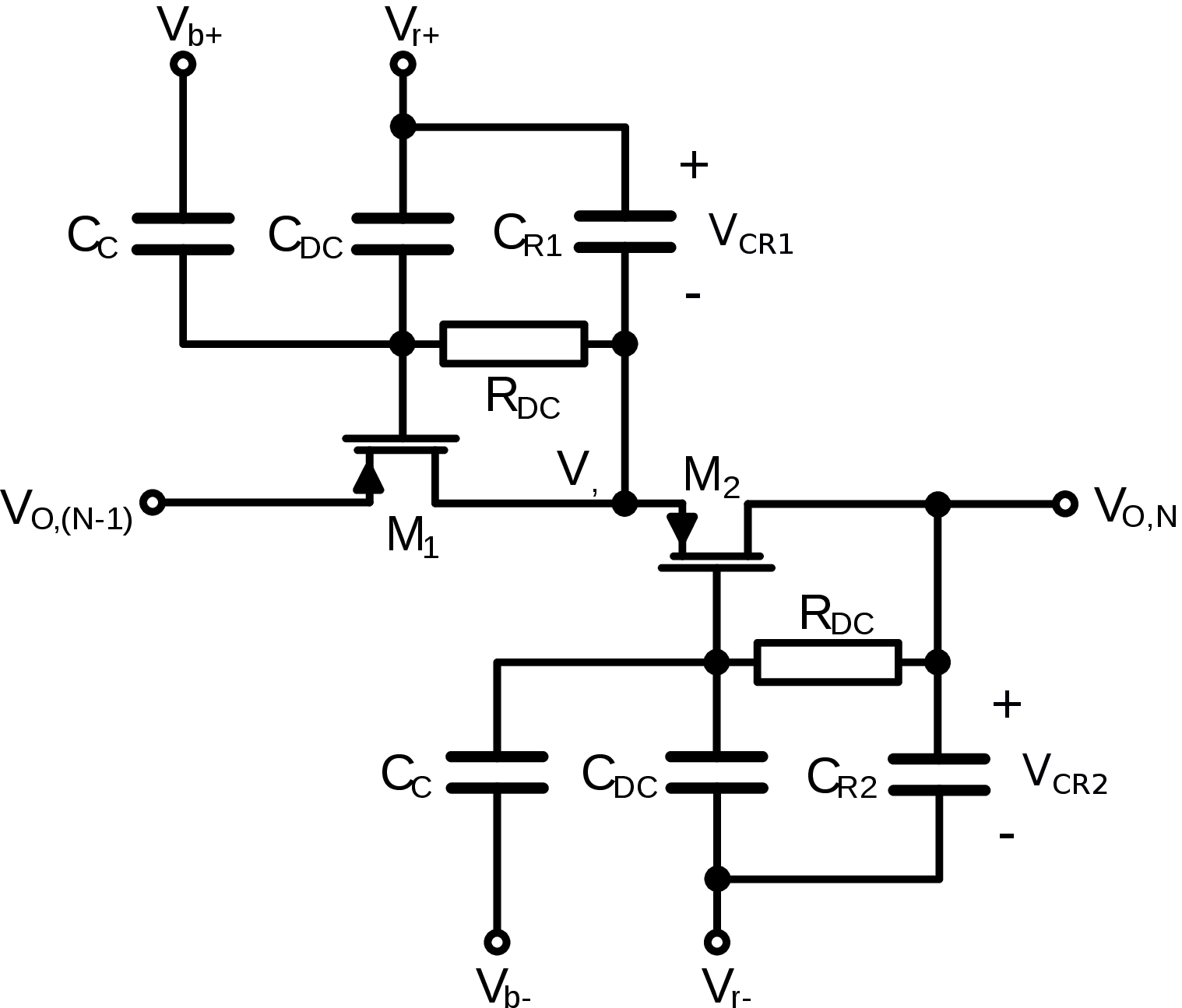,width=0.8\columnwidth}}
\label{OSCPR}
}
\end{center}
\caption{Circuit diagram of (a) boosting network and (b) a single stage of the on-chip rectifier in the RFEH~\cite{AM}.}
\label{VBNOSCPR}
\end{figure}

The RF power presented at the rectifier's input must subsequently be converted into DC power, generating a DC output voltage in an efficient manner.
The theoretical model and analysis of the rectifier designed in this work have been first presented in~\cite{AM}.
Fig.~\ref{RFEHBD} shows the block diagram of the designed RFEH that comprises a passive voltage boosting network and an orthogonally switching charge pump rectifier (OS-CPR). The circuit diagram of the boosting network and on-chip OS-CPR are shown in Fig.~\ref{VBN} and Fig.~\ref{OSCPR}.
To adequately drive the OS-CPR, the boosting network delivers large swing control ($\rm V_{b+}$ and $\rm V_{b-}$) and energy signals ($\rm V_{r+}$ and $\rm V_{r-}$). The resonant circuit of the boosting network is modeled by the self-inductance of the antenna, $\rm L_A$, its series resistance, $\rm R_A$, and capacitance $\rm C_{V,T}$ , which is the sum of the on-chip tuning capacitances ($\rm C_D$ and $\rm C_B$) and input capacitance of the rectifier ($\rm C_{R,T}$). An \hlchangeB{off-chip} inductive choke $\rm L_C$ provides a DC short at the input terminals of the rectifier to ensure a zero DC offset error at the input of the OS-CPR. In the boosting network design there is a trade-off between the value of $\rm C_{V,T}$ and $\rm L_A$. If $\rm L_A$ is made too large, to the increase voltage gain, $\rm C_{V,T}$ has to be very small. In such a case, the resonance frequency will be too sensitive to the rectifier input capacitance that changes with the load and input power. Moreover, increasing the value of $\rm L_A$ requires an inductor that is physically bigger and consequently has a bigger $\rm R_A$, which limits the voltage gain of the boosting network.

\hlchange{The rectifier circuitry (of a single stage) is made up of PMOS transistors as voltage-controlled switches ($\rm M_1$ and $\rm M_2$) and capacitors for AC coupling ($\rm C_C$) and energy storage ($\rm C_{R1}$ and $\rm C_{R2}$) \cite{AM}. 
Due to DC voltage differences within the stage, the transistors may conduct current in the backward direction in the phase they should be turned off.
Known as flow-back current this effect reduces the efficiency of the rectifier. To reduce the flow-back current, capacitors $\rm C_{DC}$ and resistors $\rm R_{DC}$ set the DC voltages $\rm V_{CR1}$ and $\rm V_{CR2}$ at the gate of $\rm M_1$ and $\rm M_2$, respectively, to guarantee that the drain and source potentials are smaller than the gate potential in the off phase.}

\hlchange{The rectifier output voltage may be too low to power a particular application.
In order to boost this voltage two techniques can be employed: more voltage-doubling stages in the rectifier or a DC-DC converter.}
Several RF energy harvesting systems reported in the literature apply DC-DC converters to up-convert the rectifier output \cite{Visser2013,Hsieh2015,Hwang2014}.
When using this approach, the DC-DC converter switching frequency can be designed to be much lower and its amplitude to be higher than that of the RF signal, resulting in a  combined efficiency of a single-stage rectifier and a DC-DC converter that can be higher than that of a rectifier with multiple stages in some cases.

The converter is then used to charge a battery or a storage capacitor.
Whenever possible, a storage capacitor is preferred due to its cost, small size and longer lifetime.

The DC-DC converter must present the optimum load to the rectifier in order to extract the maximum power out of it.
The most straight-forward way to boost the rectifier output is to use a boost converter, which most likely will be operating in Discontinuous Conduction Mode (DCM) due to the low load current, limited by the low available power.
However, the average input resistance of the boost converter is dependent on the output voltage:
\begin{equation}
R_{in} = \frac{2 L}{D^2 T} \left( 1 - \frac{V_{in}}{V_{out}} \right),
\end{equation}
in which $T$ is the switching \textcolor{black} {period}, $D$ is the duty cycle, $L$ is the inductor value, $V_{in}$ is the rectifier output voltage (input of the DC-DC converter) and $V_{out}$ is the boost converter output voltage.
While charging a storage capacitor, $V_{out}$ will increase every cycle, taking the rectifier load away from its optimal value.
This problem can be solved by employing a buck-boost converter, which isolates the input from the output while operating in DCM and which presents the following average input resistance \cite{Martins2016}:
\begin{equation}
R_{in} = \frac{2 L}{D^2 T}.
\label{eq:bbcrin}
\end{equation}

\begin{figure}
\centering
\includegraphics[width=\linewidth]{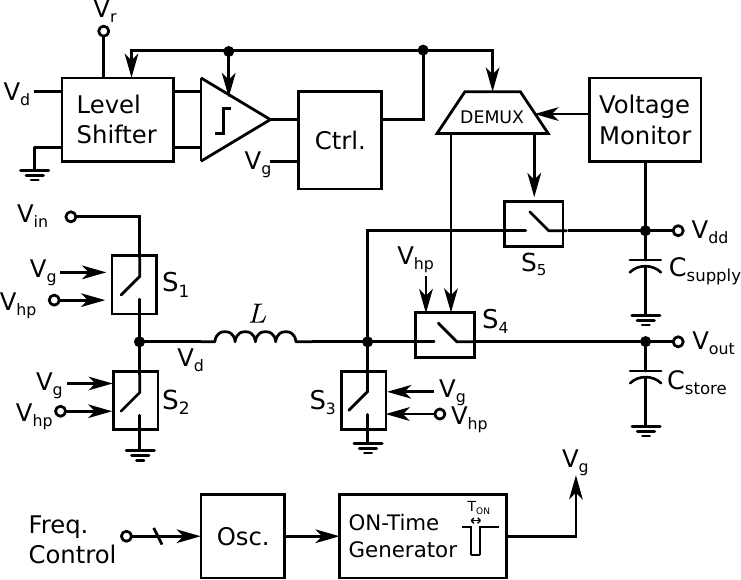}
\caption{Buck-boost converter core circuit.}
\label{fig:bbconv}
\end{figure}

\hlchange{
In this work we propose a non-inverting buck-boost converter to serve as the DC-DC converter. 
The core of the converter is formed by switches $S_1$-$S_4$, inductor $L$ and storage capacitor $C_{store}$ as depicted in Fig. \ref{fig:bbconv}.
Ultimately, the goal of the energy harvesting front-end is to charge $C_{store}$, which will then be used to power the sensor node's circuits.
Capacitor $C_{supply}$ stores the energy necessary to operate the energy harvesting circuits, \textcolor{black}{i.e., it provides supply voltage $V_{DD}$ to them}.
The start-up circuit charges capacitor $C_{supply}$.
It consists of a charge pump and a ring oscillator that can operate from a low input voltage, but it has a low efficiency.
\hlchange{The start-up charge pump is turned off by the voltage monitor when $C_{store}$ has enough voltage so that the buck-boost converter can operate and charge $C_{store}$ by itself.}
When the voltage on $C_{supply}$ reaches its maximum value the voltage monitor redirects the switching signal $V_S$ to switch $S_4$ and keeps $S_5$ off, to start charging $C_{store}$.
When $V_{DD}$ drops below a certain lower limit, $S_5$ will be switching while $S_4$ will be off and the converter will charge $C_{supply}$.
Capacitor $C_{rec}$ must be large enough so that its voltage ripple due to the inductor current is negligible.
Not depicted in the block diagram are the Zero Current Detection (ZCD) circuit, which switches off $S_5$ or $S_4$ when the inductor current drops to zero, the oscillator and the ON-time generator \cite{Martins2016}.}

\begin{figure*}
\centering
\subfigure[]{\includegraphics{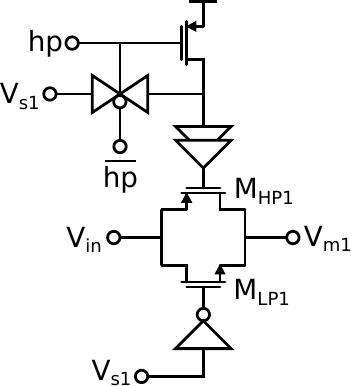}}
\subfigure[]{\includegraphics{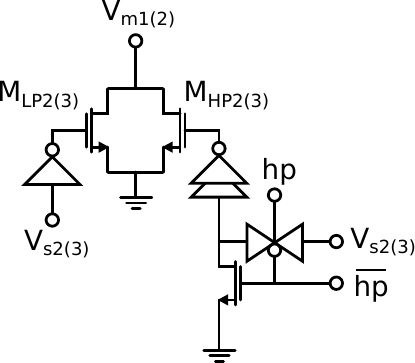}}
\subfigure[]{\includegraphics{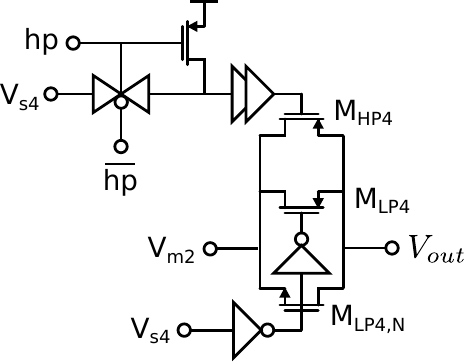}}
\subfigure[]{\includegraphics{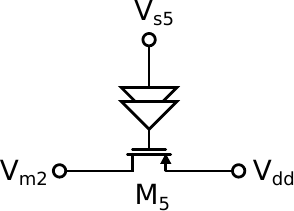}}
\caption{Switches circuit schematics: (a) switch $S_1$, (b) switches $S_2$ and $S_3$, (c) switch $S_4$ and (d) switch $S_5$.}
\label{fig:switches}
\end{figure*}

\hlchange{
When the input/output power decreases, the converter efficiency reduces and the switching loss becomes one of the dominant energy loss factors \cite{Hu2011}. 
This loss can be reduced by decreasing the switching frequency and increasing the inductor value, which will increase the size of the device.
The switching loss can be further reduced by using smaller switches, but there exists a trade-off with conduction loss, since the ON resitance of the switch will increase as its width decreases.
When the input/output power increases the conduction loss becomes dominant.
To design a converter that is efficient at both, e.g., $1 \unit{\mu W}$ and $1 \unit{mW}$ input power, i.e., over a range of 3 decades, switches $S_1$-$S_4$ can be configured to operate in either low-power or high-power mode.
Fig. \ref{fig:switches} presents the circuit diagrams of the switches.
The signal $hp$ controls in what mode the switches will operate.
In the low-power mode, only the $M_{LP}$ transistors are switching while the $M_{HP}$ transistors are always turned off.
In the high-power mode, both $M_{LP}$ and $M_{HP}$ are switching.
Therefore, in the low-power mode, the power necessary to drive the switches is reduced and in the high-power mode the switches series resistance is reduced.
Because for low input power the voltage $V_{in}$ is also low, in switch $S_1$ the low-power transistor $M_{LP1}$ is an NMOS.
In switch $S_4$ an extra NMOS is used ($M_{LP4,N}$) in order to increase the efficiency when the output voltage is low.
However, most of the energy transfer happens when $V_{out}$ is high ($E = CV^2/2$), so we use only one NMOS and keep the added parasitic capacitance at node $V_{m2}$ low. 
Switch $S_5$ consists of a single transistor, because the charging of $C_{supply}$ is very short and does not have a strong influence on the efficiency.}

In the ZCD circuit, the voltage at node $V_{m1}$ is compared to the ground voltage to detect when the inductor current crosses zero.
The comparator operates during a brief period after switches $S_1$ and $S_3$ are turned off.
After the current crosses zero, the comparator is switched off again.
This reduces the average power consumption, but it can be further decreased, while maintaining its speed,  by applying dynamic biasing to the comparator \cite{Martins2016}.

The switching frequency of the buck-boost converter can be selected dynamically, so its average input resistance can change (according to (\ref{eq:bbcrin})). 
\textcolor{black}{Here, the MPPT circuit performs this frequency selection in order to provide the best load to the rectifier.}

\begin{figure}
\centering
\includegraphics[width=\linewidth]{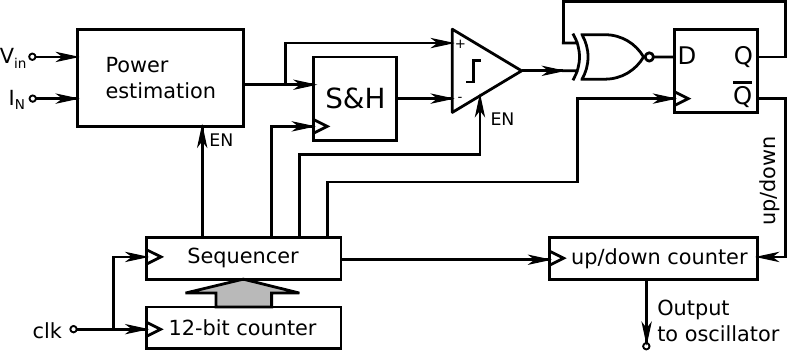}
\caption{MPPT block diagram.}
\label{fig:mpptblockdiag}
\end{figure}

The rectifier is usually the least efficient block in the power conversion chain \cite{Visser2013}.
\hlchange{Therefore, by maximizing its output power, taking it alway from a low efficiency condition to a high efficiency one, we optimize the entire power conversion chain efficiency.}
As mentioned before, one of the steps for doing so is optimizing the rectifier load, which is the average input resistance of the DC-DC converter.
We choose to base the MPPT circuit designed on the Perturb and Observe algorithm, due to its inherent low-power consumption \textcolor{black}{\cite{Bui2016}}.
The MPPT block diagram is presented in Fig. \ref{fig:mpptblockdiag}.

\begin{figure}
\centering
\includegraphics[width=\linewidth]{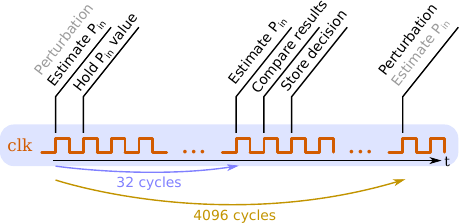}
\caption{Timing diagram of the MPPT circuit.}
\label{fig:timingdiagram}
\end{figure}

The sequence of events that results in maximum power tracking is presented in Fig. \ref{fig:timingdiagram}.
At first, the rectifier output power is estimated and held in the sample and hold (S\&H) block.
A perturbation is applied, i.e., the oscillator frequency is either increased or decreased (depending on the D flip-flop output).
After $32$ clock cycles the output power is estimated once again and it is compared to the previous value.
If the result of the comparison is positive (i.e., the power increased due to the perturbation) the value stored by the flip-flop remains unchanged, otherwise it is inverted.
This value is fed to the up/down counter, which is activated to introduce the perturbation and the output of which controls the oscillator bias current.
\hlchange{The analog circuits are turned off when not in use and after a long time (4096 clock cycles from the start, in this case) the procedure repeats itself.}

The MPPT must dissipate very little power in order to have as little influence on the total power loss as possible.
This can be achieved with a low sampling rate, turning the circuits off when not in use.
However, it would require a sample and hold circuit that can hold for a very long time (which dissipates power).
\hlchange{Instead, we choose to sample the rectifier output power one extra time, within a shorter period ($32$ clock cycles, in this case).}
The number of cycles between the two power estimations is selected to provide enough settling time to the rectifier output capacitor.

The power estimation itself is based on the equation of the input power of a buck-boost converter in DCM:
\begin{equation}
P_{in} = \frac{V_{in}^2}{R_{in}} = V_{in}^2 \frac{D^2 T}{2 L} = V_{in}^2 f_s \frac{T_{ON}^2}{2 L}.
\end{equation}
The switching frequency $f_s$ is proportional to the oscillator bias current $I_B$, which leads to:
\begin{equation}
P_{in} \propto V_{in}^2 I_B.
\label{eq:ppin}
\end{equation}
Knowing that the other factors are constant, we just have to maxime $V_{in}^2 I_B$ to maximize the input power.
The same result is obtained if we maximize the square root of this value, which can be readily obtained using a differential pair in strong inversion.

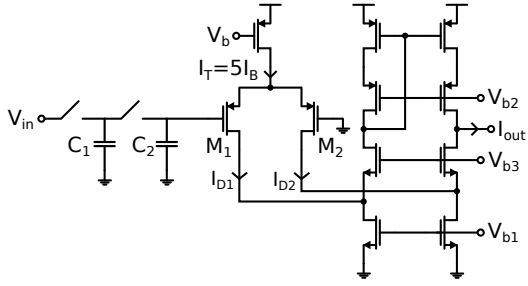
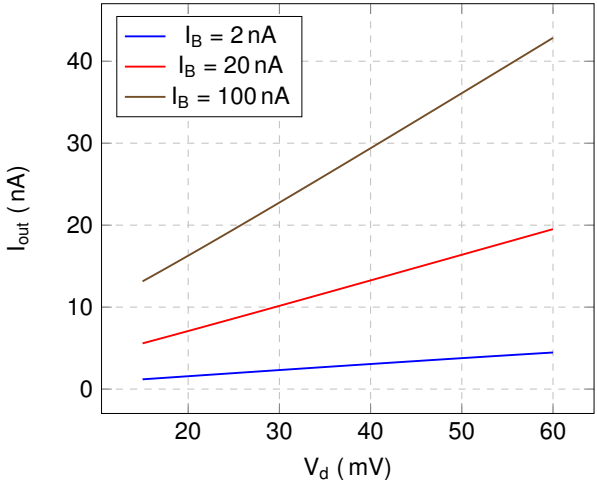
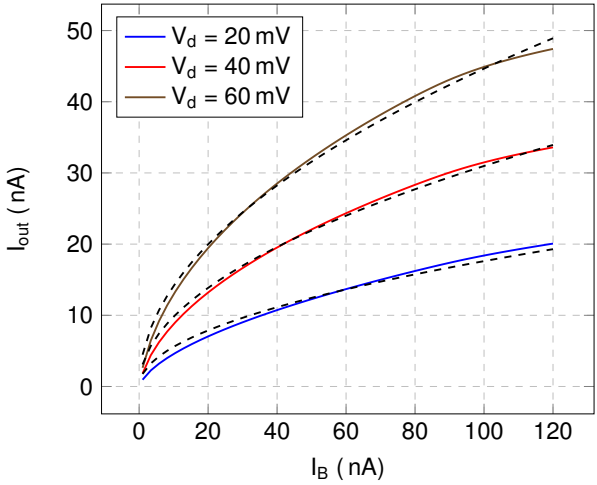
\begin{figure}
\centering
\subfigure[]{
\includegraphics[width=\linewidth]{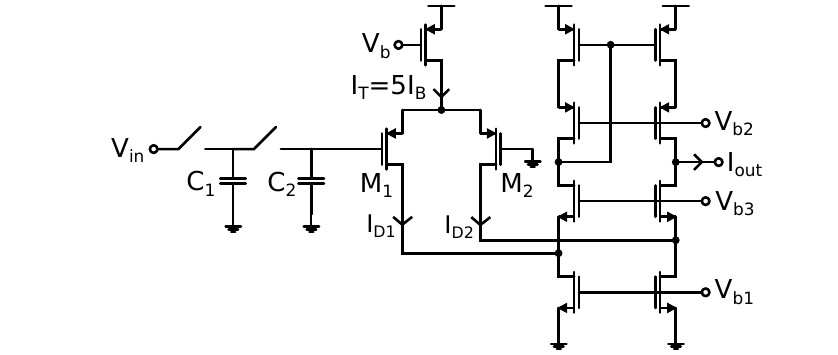}
\label{fig:pinest}
}
\subfigure[]{
	\begin{tikzpicture}[scale=0.9, trim axis left]
	\begin{axis}[xlabel={$\mathrm{V_{d}}$ ($\unit{mV}$)}, ylabel={$\mathrm{I_{out}}$ ($\unit{nA}$)}, grid=major, table/col sep=comma, no markers, legend pos=north west,width=\linewidth]
	\addplot+[thick] table[x expr=\thisrowno{0}*1e3, y expr=\thisrowno{1}*1e9] {Power/data/ioutvin2n.csv};
	\addplot+[thick] table[x expr=\thisrowno{0}*1e3, y expr=\thisrowno{1}*1e9] {Power/data/ioutvin20n.csv};
	\addplot+[thick] table[x expr=\thisrowno{0}*1e3, y expr=\thisrowno{1}*1e9] {Power/data/ioutvin100n.csv};
	\legend{$\mathrm{I_B} = 2 \unit{nA}$,$\mathrm{I_B} = 20 \unit{nA}$,$\mathrm{I_B} = 100 \unit{nA}$};
	\end{axis}
	\end{tikzpicture}
\label{fig:pinestv}
}
\subfigure[]{
	\begin{tikzpicture}[scale=0.9, trim axis left]
	\begin{axis}[xlabel={$\mathrm{I_{B}}$ ($\unit{nA}$)}, ylabel={$\mathrm{I_{out}}$ ($\unit{nA}$)}, grid=major, table/col sep=comma, no markers, legend pos=north west,width=\linewidth]
	\addplot+[thick] table[x index=0, y index=1, x expr=\thisrowno{0}*1e9, y expr=\thisrowno{1}*1e9] {Power/data/Iout.csv};
	\addplot+[thick] table[x index=0, y index=3, x expr=\thisrowno{0}*1e9, y expr=\thisrowno{3}*1e9] {Power/data/Iout.csv};
	\addplot+[thick] table[x index=0, y index=5, x expr=\thisrowno{0}*1e9, y expr=\thisrowno{5}*1e9] {Power/data/Iout.csv};
	\addplot+[thick,dashed,black] table[x expr=\thisrowno{0}*1e9, y expr=\thisrowno{1}*1e9] {Power/data/sqrt20m.csv};
	\addplot+[thick,dashed,black] table[x expr=\thisrowno{0}*1e9, y expr=\thisrowno{1}*1e9] {Power/data/sqrt40m.csv};
	\addplot+[thick,dashed,black] table[x expr=\thisrowno{0}*1e9, y expr=\thisrowno{1}*1e9] {Power/data/sqrt60m.csv};
	\legend{$\mathrm{V_d} = 20 \unit{mV}$,$\mathrm{V_d} = 40 \unit{mV}$,$\mathrm{V_d} = 60 \unit{mV}$}
	\end{axis}
	\end{tikzpicture}
\label{fig:pinesti}
}
\caption{Rectifier output power estimator: (a) schematic; (b) output current versus differential input voltage; (c) output current versus $I_B$ (dashed lines represent the best fitting square root curves).}
\end{figure}

The circuit employed to do the power estimation is presented in Fig. \ref{fig:pinest}.
The difference between the drain currents in the differential pair is given by:
\begin{equation}
I_{D1} - I_{D2} = \sqrt{2 K} \sqrt{I_T} V_d,
\end{equation}
in which $K = \frac{1}{2}\mu_n C_{ox} \frac{W}{L}$, $V_d$ is the differential input, which is a fraction of the input voltage, and $I_T$ is the tail current, which is proportional to $I_B$.
Therefore, the output current of this circuit is proportional to the square root of \eqref{eq:ppin} and maximizing it will maximize the rectifier output power.

The circuit topology shown in Fig. \ref{fig:pinest} was chosen because of the limited voltage headroom that the differential pair must operate in, recalling that the minimum $V_{DD}$ for which this circuit must operate is $1.2 \unit{V}$ (because it shares the same supply as the buck-boost converter) and that the differential pair must be in strong inversion.
In the simulation results, we can observe the linear variation with $V_d$, Fig. \ref{fig:pinestv}, and the square root variation with $I_B$, Fig. \ref{fig:pinesti}, in which the dashed lines are the best fitting square roots.

The current output of the power estimator is fed into a diode connected NMOS, which converts the current into a voltage.
The sample and hold circuit consists of a simple switch and capacitor to hold the voltage value.
The comparator employed is a StrongARM comparator \cite{Kobayashi1992}.
The other blocks of the MPPT are all digital: the 12-bit counter provides the input to the sequencer, which enables/disables and generates the clock signal for all the other blocks; the up/down counter sets the oscillator frequency.
\label{sec:ehmeth}
\hlchange{
\subsection{Implementation and Results}}
The rectifier presented here has been implemented in silicon using AMS 0.18$\rm\mu$m CMOS IC technology. 
In order to select the operating frequency of the rectifier, we analyzed the rectifier \hlchangeB{Power Conversion Efficiency (PCE)} in three different ISM bands: $13.56$, $433$ and $915 \unit{MHz}$.
The low frequency $13.56 \unit{MHz}$ ISM band, compared to the others, presents better performance since at high frequencies the parasitic capacitances of the transistors add significant losses \cite{AM}.
Moreover, at this low frequency more power can be radiated from the RF power source ~\cite{website}.
\hlchange{Table~\ref{devices} shows the component values of the designed RF energy harvester which has a number of stages, $N$ equal to $5$.}
For all the measurement results presented in this subsection, the RF power source at $13.56 \unit{MHz}$ is calibrated for a distance of $10 \unit{cm}$ (coupling factor of $0.004$) between the antenna of the RF source and the antenna of the RF energy harvester.

\begin{table}
\footnotesize
\renewcommand{\arraystretch}{1.5}
\caption{RFEH Component Values}
\label{devices}
\centering
\begin{tabular}{c|c||c|c}
\hline
Device & Value & Device & Value \\
\hline
$\rm C_B$ & $7.5 \unit{pF}$ & $\rm C_{DC}$ &$\rm\simeq 90 \unit{fF}$ \\
\hline
$\rm C_D$ & $19.5 \unit{pF}$ & $\rm R_{DC}$ & $350 \unit{k\Omega}$ \\
\hline
$\rm C_{R,T}$ & $\rm\simeq 17 \unit{pF}$ & $\rm C_{R1},C_{R2}$ & $9.7 \unit{pF}$\\
\hline
$\rm C_C$ & $9 \unit{pF}$ & $\rm M_1,M_2$ & $750 \unit{\mu m}$/$0.2 \unit{\mu m}$\\
\hline
\end{tabular}
\end{table}

\hlchange{Before presenting the measurement results, a new analysis of the PCE of the rectifier is presented and compared with the state of the art EHs for IoT applications shown in Table \ref{tab:ehpwr}. }
First, three possible definitions of power efficiency are given and then discussed and compared.
The measurements of the rectifier presented in the previous subsection are discussed.

The power conversion efficiency is the ratio between the power delivered to the load and the input power. Although the PCE definition is very clear, the input power can be defined in several ways.
We recognize three definitions of input power to present PCE measurement results.

The first definition is the \textit{theoretical input power} ($\rm P_{INtheor.}$), which is the input power defined as $\rm P_{INtheor.}= V_A^2/(2{\rm Re}\{Z_A\})$.
The second definition is the \textit{measured input power at the antenna} ($\rm P_{INant.}$).
The third definition is the \textit{estimated input power at the rectifier circuit} ($\rm P_{IN}$).
Most of the references on RF energy harvesting present the power conversion efficiency using $\rm P_{IN}$ as input power definition.
Comparing the three definitions, PCE will be the lowest for the $\rm P_{INtheor.}$ since it does not take into account losses in the antenna and loading effects.
The highest PCE is seen for an input power $\rm P_{IN}$ that takes into account all the losses in front of the input of the rectifier. Therefore $\rm P_{IN}$ is smaller than $\rm P_{INant.}$ and consequently PCE is bigger.

\begin{figure}
\centering
    \centering
    \includegraphics[width=0.9\columnwidth]{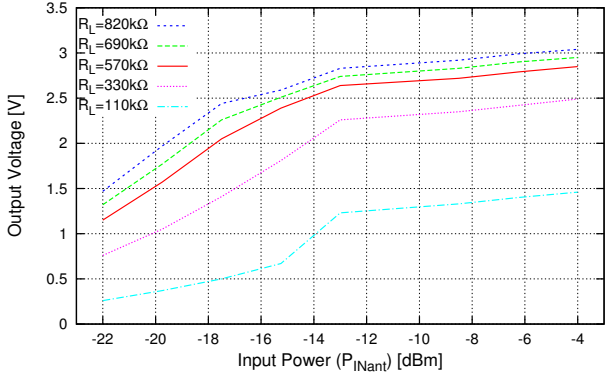}
    \caption{Measured output voltage of the RFEH as a function of input power for $110 \unit{k\Omega} \leq R_L \leq 820 \unit{k\Omega}$}
    \label{fig:Vopi}
\end{figure}

\begin{figure}
    \centering
    \includegraphics[width=0.9\columnwidth]{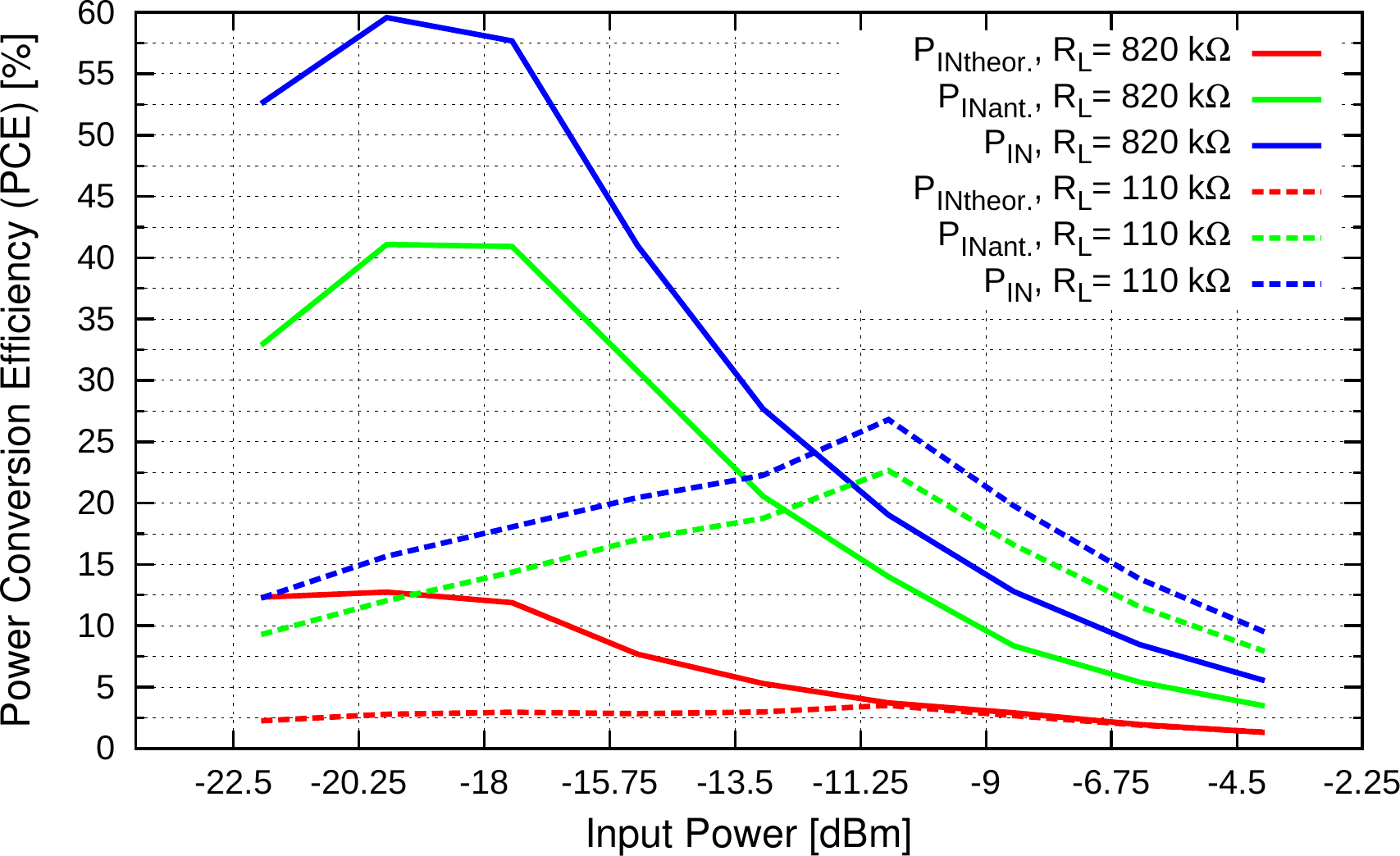}
    \caption{Measured power conversion efficiency of the RFEH as a function of input power for $R_L = 110 \unit{k\Omega}$ (dashed lines) and $820 \unit{k\Omega}$ (solid lines)}
    \label{fig:eff110}
\end{figure}

\begin{figure}
    \centering
    \includegraphics[width=0.9\columnwidth]{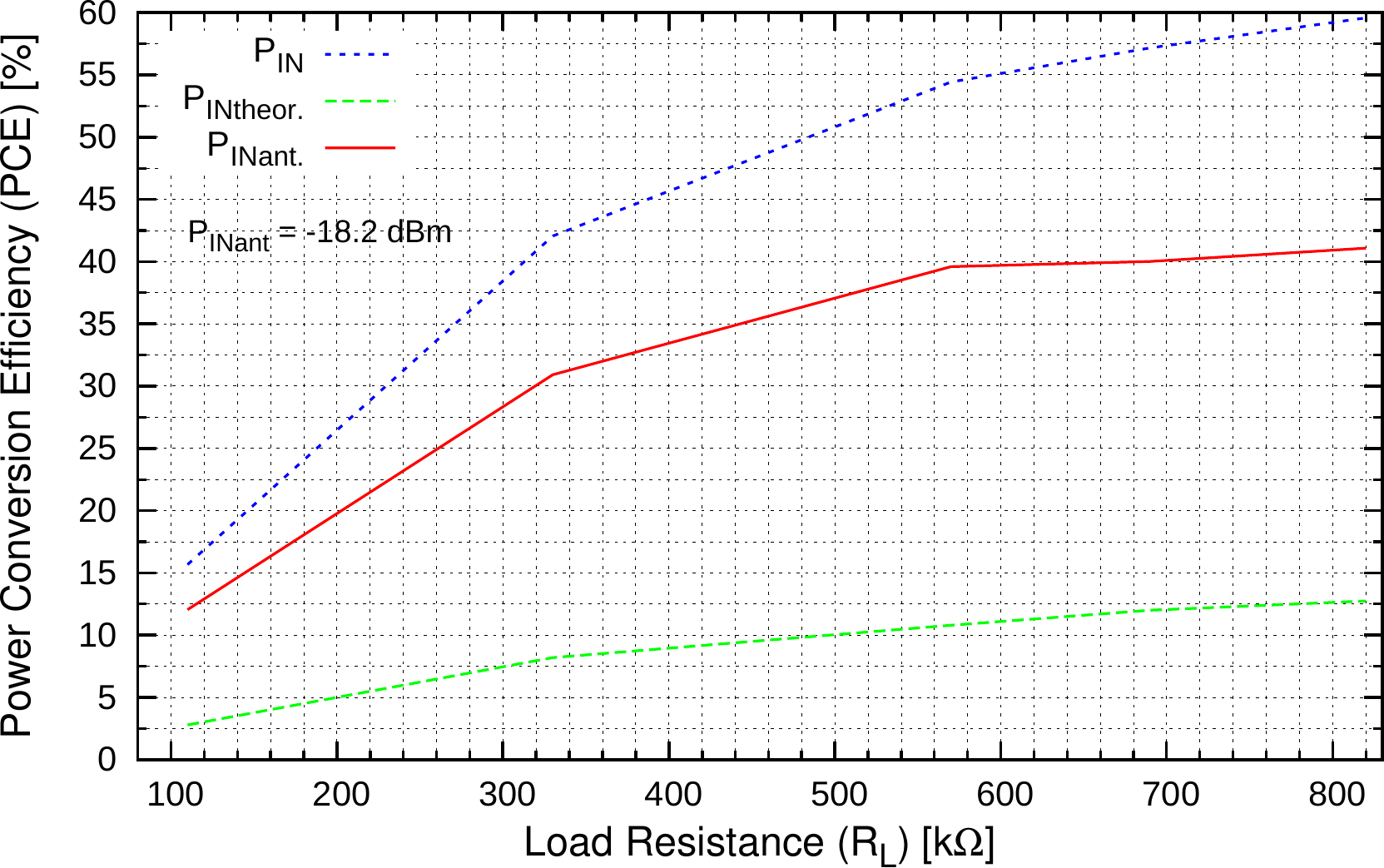}%
    \caption{Measured power conversion efficiency of the RFEH as a function of $\rm R_L$ for $\rm P_{IN} = -18 \unit{dBm}$}
    \label{fig:effRL}
\end{figure}

Fig.~\ref{fig:Vopi} presents the measured output voltage of the RFEH as a function of $\rm P_{INant}$ for $\rm 100 \unit{k\Omega}\leq R_L\leq850\unit{k\Omega}$.
From Fig.~\ref{fig:Vopi} it can be noticed that the output voltage increases with input power and $\rm R_L$, which means that a system powered by the RFEH has to operate from very little power, otherwise the sensitivity (minimum input power required for system operation) to the RF source is degraded. 
Fig.~\ref{fig:eff110} and Fig.~\ref{fig:effRL} show the PCE behavior as a function of input power for different loads.
\hlchangeB{
The value of $P_{in}$ is estimated by measuring the input impedance of the rectifier using a VNA.
The presented measurements were performed employing a VNA, a function generator and a multimeter.}

In Table \ref{tab:ehpwr}, some energy harvesters already published in the literature have been listed. The input available power heavily depends on the source of energy used.   Therefore, when comparing this EH with the work in Table \ref{tab:ehpwr}, not only the peak power efficiency should be considered, but also the input power range across which the EH can operate plays an important role. 
The input power range of the EH presented is from $-20 \unit{dBm}$ \hlchangeB{($10 \unit{\mu W}$)} to $-4 \unit{dBm}$ \hlchangeB{($400 \unit{\mu W}$)} with a maximum PCE of $60 \%$ achieved at $P_{IN}=- 20 \unit{dBm}$. This important achievement allows to pull up the curve of the output power of an energy harvester depicted in Fig. \ref{power_scenario}.

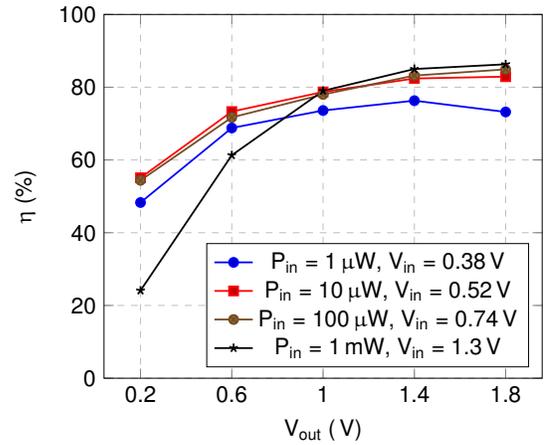
\begin{figure}
\centering
	\begin{tikzpicture}[scale=0.85]
	\begin{axis}[xlabel={$\mathrm{V_{out}}$ ($\unit{V}$)}, ylabel={$\eta$ ($\pct$)}, grid=major, ymin=0, ymax=100, legend pos=south east,
			xtick={0.2, 0.6, 1.0, 1.4, 1.8}]
	\addplot+[thick] coordinates { 
		(0.2, 48.3)
		(0.6, 68.8)
		(1.0, 73.6)
		(1.4, 76.3)
		(1.8, 73.2)
		}; 
	\addplot+[thick] coordinates { 
		(0.2, 55.1)
		(0.6, 73.3)
		(1.0, 78.7)
		(1.4, 82.4)
		(1.8, 82.9)
		}; 
	\addplot+[thick] coordinates { 
		(0.2, 54.4)
		(0.6, 71.7)
		(1.0, 78.0)
		(1.4, 83.2)
		(1.8, 84.9)
		}; 
	\addplot+[thick] coordinates { 
		(0.2, 24.1)
		(0.6, 61.3)
		(1.0, 79.0)
		(1.4, 85.0)
		(1.8, 86.3)
		}; 
	\legend{{$\mathrm{P_{in}}=1 \unit{\mu W}$, $\mathrm{V_{in}}=0.38 \unit{V}$},
		{$\mathrm{P_{in}}=10 \unit{\mu W}$, $\mathrm{V_{in}}=0.52 \unit{V}$},
		{$\mathrm{P_{in}}=100 \unit{\mu W}$, $\mathrm{V_{in}}=0.74 \unit{V}$},
		{$\mathrm{P_{in}}=1 \unit{mW}$, $\mathrm{V_{in}}=1.3 \unit{V}$}}
	\end{axis}
	\end{tikzpicture}
\caption{Efficiency results of the buck-boost converter for various output voltages. \cite{Martins2016}}
\label{fig:res}
\end{figure}

\hlchange{
Regarding the DC-DC converter, it was designed to operate with input powers ranging from $1 \unit{\mu W}$ to $1 \unit{mW}$ and input voltages ranging from $0.38$ to $1.3 \unit{V}$.
As commentend before, the switching losses for low-power levels become dominant and a reduction of the switching frequency is necessary.}
To avoid increasing the conduction losses to high levels, a large inductor value must be employed, which will increase the size of the device.
Because off-the-shelf inductors with larger inductances are bulky, and in order to keep the system size as small as possible, in this work we have selected a $220$-$\unit{\mu H}$ power inductor with a parasitic series DC resistance of $21.1 \unit{\Omega}$ (Coilcraft XPL2010-224ML).
\hlchangeB{The capacitors $C_{rec}$, $C_{store}$ and $C_{supply}$ are also external components and their values are $8.5 \unit{nF}$, $22 \unit{\mu F}$ and $20 \unit{nF}$, respectively. The value of $C_{store}$ and $C_{supply}$ can be increased. However, the additional leakage will affect the efficiency.}
The simulated efficiency (in AMS $0.18 \unit{\mu m}$ CMOS technology) of the buck-boost converter for various output voltages is presented in Fig. \ref{fig:res}.
The peak efficiency of the buck-boost converter is $76.3 \%$ at an input power of $1 \unit{\mu W}$ and $86.3 \%$ at $1 \unit{mW}$.

Circuit simulations show that the implemented MPPT circuit dissipates $17.4 \unit{nW}$ from a $1.8$-$\unit{V}$ supply.
This simulation was performed for an operating frequency of $20 \unit{kHz}$ (oscillator biased with $2 \unit{nA}$), which is the configuration to achieve $1 \unit{\mu W}$ of input power.
If the frequency and hence the oscillator biasing current increase, the power consumption increases as well.
At $1 \unit{MHz}$, which is the maximum frequency of the system and the one in which it consumes the highest power, the power consumption of the MPPT is $278.5 \unit{nW}$.

The combination of the DC-DC converter, presented previously, with the MPPT allows to boost the rectifier voltage and charge a storage capacitor while presenting the best load to the rectifier, thereby optimizing its efficiency.
Moreover, the MPPT circuit power consumption is low, allowing for efficient harvesting down to $1 \unit{\mu W}$ available input power.

\hlchange{
\subsection{Discussion}
The rectifier was designed for the operating frequency of $13.56 \unit{MHz}$.
A higher frequency will result in an increase of power loss, which can be reduced by using a more advanced technology that features a shorter transistor length that will, in turn, result in a lower parasitic capacitance and a lower ON resistance.
This, however, may result in a larger flow-back current.
So, even in more advanced technologies, a balance between power loss provided due to flow-back and to conduction losses must be found.
The designer must select the transistor dimensions that fit best.

One can see that different PCEs are achieved for different loads and input power levels, which indicates that the design of the RFEH strongly depends both on its output power and on the power received by the antenna. The latter may change due to different coupling factors as a result of variations in distance and/or alignment of antennas.
So it is important to match the load to the varying power levels presented to the rectifier.
This task is performed by the presented DC-DC converter and MPPT circuits, which are designed for a specific input power range.
If one wishes to increase the power that can be processed by the DC-DC converter, and keep it in the DCM operating mode as discussed previously, one must decrease the inductor size to achieve higher currents and possibly increase the switching frequency.
To be able to harvest at lower input power levels, the frequency must be further reduced and the inductor increased, which will eventually require a bulky inductor, if the \hlchangeB{Effective Series Resistance (ESR)} is to be kept low.

}

\section{Receiver: co-design of low-noise amplifier and antenna}
\label{sec:datarx}
As addressed in \cite{Mark_TCAS2}, the co-design principle presented in Section \ref{sec:ehmeth} also holds for the interface between an LNA and an antenna.
Here we demonstrate the \hlchangeB{Noise Figure (NF)} improvement introduced by the proposed co-design principle.

\begin{figure}
\centering{}\includegraphics[width=0.6\columnwidth]{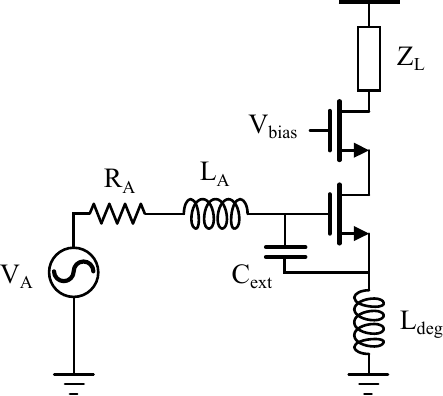}\caption{\label{fig:Model-of-an}Interface model of an inductively degenerated
CMOS LNA directly connected to an inductive antenna impedance.}
\end{figure}

\hlchange{
\subsection{Circuit Description}}
The co-design of any antenna-electronics interface starts by optimizing
the load impedance, which in this example is \hlchangeB{the input impedance of a} narrowband
LNA. The well-known inductively degenerated CMOS cascode LNA topology
\cite{LNA} is used as it provides an easy way of adjusting the LNA
input impedance. The LNA is directly connected to an inductive antenna
as depicted in Fig. \ref{fig:Model-of-an}. The information is sensed
with a CMOS gate, meaning that voltage is the preferred signal quantity
to maximize. 

For this particular LNA implementation, the interface impedance is
defined as $Z_{int}=R_{A}+j\omega(L_{A}+L_{deg})=R_{A}+jX_{A}$
as the total inductance in the interface is the sum of the antenna
and the degeneration inductors. If the interface is conjugate matched,
the antenna load voltage can be approximated by (\ref{eq:VoutBoost}) for large values of Q. The minimum noise factor for low and
medium frequencies can be approximated as:

\begin{equation}
F_{min}\approx1+\delta\frac{R_{g}}{R_{A}}+\underset{co-design}{\underbrace{\frac{R_{A}}{X_{A}^{2}}}}\underset{LNA}{\underbrace{\left(\frac{\gamma}{g_{m}}+\frac{4}{g_{m}^{2}R_{L}}\right)}}\label{eq:NF}
\end{equation}
Here, $g_{m}$ denotes the transconductance of the MOS
transistor, $R_{g}$ is the transistor gate resistance
and $R_{L}$ is the equivalent thermal noise resistance
of the LNA's load and subsequent stages. The coefficient $\gamma$
is often between 2/3 and 2, depending on the transistor size and the
technology. 
Notice that the 'LNA' term in (\ref{eq:NF}) only depends on the LNA
circuit parameters and can be minimized by increasing the MOS transistor's
bias current and gate area. The 'co-design' term allows to reduce
the noise factor without additional power consumption by using a high-Q
impedance interface  \cite{Mark_TCAS2}.

\hlchange{
\subsection{Implementation and Results}}
As a proof of concept, a narrow band LNA with a center frequency of $900 \unit{MHz} $ is designed to be implemented in AMS 0.18 um technology and its design parameters
are kept constant during the circuit simulations ($g_{m}=366
\unit{\mu S}$, $C_{gs}=4 \unit{fF}$, $R_{g}=18 \unit{\Omega}$,
$R_{L}=10 \unit{k\Omega}$, $\gamma=1.1$). The
LNA input impedance is varied by tuning $L_{deg}$ and
$C_{ext}$ while the antenna impedance is subsequently
conjugate matched to the LNA input for each case. 
The difference
in noise factor is thus only determined by the difference in interface
impedance, when considering ideal antenna impedance and matching components . 

\begin{figure}
\centering{}\includegraphics[width=0.9\columnwidth]{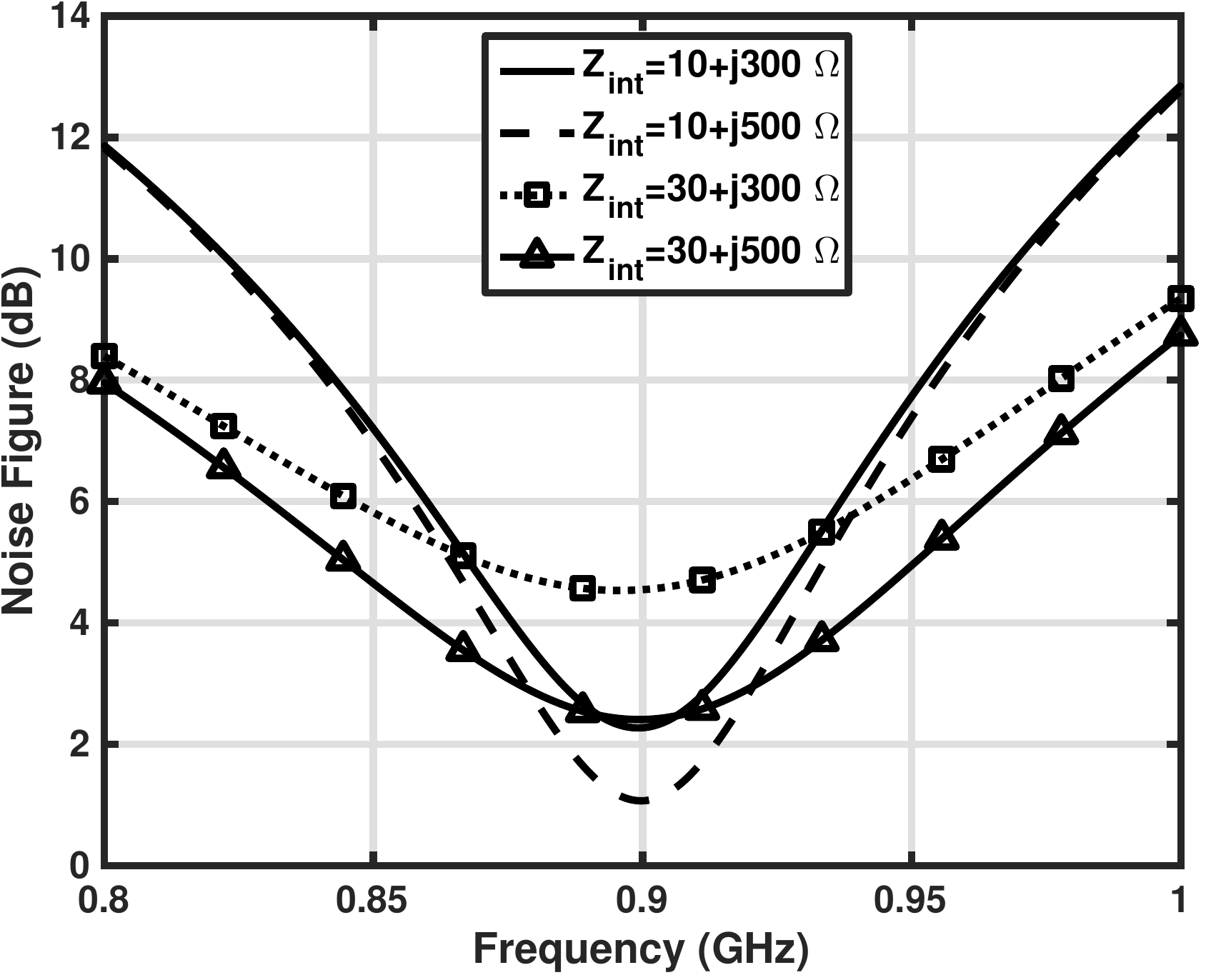}\caption{\label{fig:Simulated-Noise-Figure}Simulated narrowband LNA noise
figure for various interface impedances \cite{Mark_TCAS2}.}
\end{figure}

The impact of the 'co-design' term can be confirmed by the simulated
NF for various interface impedances, as shown
in Fig. \ref{fig:Simulated-Noise-Figure}. Note that, in order to
clearly demonstrate the impact of the interface impedance $Z_{int}$,
some of $Z_{int}$ in Fig. \ref{fig:Simulated-Noise-Figure}
have a big reactance and hence require an impractical value of $L_{deg}$
(e.g., 50 nH). 
\hlchange{In practice, the feasibility of the required $L_{deg}$
and other design constraints may result in interface impedances that
are different from those in Fig. \ref{fig:Simulated-Noise-Figure}.
\textcolor{black}{Equation} (\ref{eq:NF}) and Fig. \ref{fig:Simulated-Noise-Figure} suggest
that $R_{A}$ should be as low and $X_{A}$
as high as possible to reduce NF. In practice however, this will cause
the antenna radiation efficiency to drop considerably when the antenna
conduction loss resistance becomes comparable to $R_{A}$
\cite{Wearable}. In this case, a minimum $R_{A}$ should
be selected during the optimization process. The LNA input, however,
can be designed for maximum parallel resistance (i.e., purely capacitive
input impedance) and therefore would increase the load voltage by
6 dB when keeping $R_{A}$ fixed at the minimum value \cite{Nauta}.}

\hlchange{
\subsection{Discussion}
From Eq. \ref{eq:NF}, it is worth to notice that for a given minimum noise factor, the co-design principle allows to reduce the $g_m$ of the MOS transistor, and therefore the power consumption of the LNA can be reduced.
A limitation of this principle is imposed by the fact that a purely capacitive input impedance in theory would increase the voltage even further, but in this case would require a purely inductive antenna with infinitely small antenna radiation resistance and conduction loss resistance, which of course is not realizable.

The co-design technique presented in this section is specially useful for the scenarios d) and e), in which the receiver has its own antenna, because the antenna impedance can be selected to match the impedance required by the receiver.
However, if the transmitter and receiver need to share the same antenna, they can be designed to have the same impedance, increasing the number of system design iterations to be performed.

}

\section{Low-Power Data transmission}
\label{sec:datatx}
Once powered by the RFEH, the system can read data from the sensors and transmit it.
\hlchange{Active data transmission is usually employed to increase the range over which the sensor node can operate, but with the drawback of consuming more power.
With the promise to offer both low-power operation and a high channel capacity, research efforts have been concentrated on the development of transmitters and receivers for UWB communication.
There are two permitted FCC unlicensed bands: sub-GHz (up to 950 MHz) and 3.1-10.6 GHz.
A sub-GHz UWB signal can reach a longer distance if compared to a $3$-$10$ \unit{GHz} one, with the same transmitted power, due to the lower free space loss.
However, a sub-GHz transmitter has to comply with a steep roll-off at 950 \unit{MHz} \cite{FCC_reference}, which poses a design challenge. In this section we present a novel low-power sub-GHz UWB transmitter (LPUT) topology.}

\hlchange{
\subsection{Circuit description}}
\label{transmitter_method}

\begin{figure}
\centerline{\psfig{figure=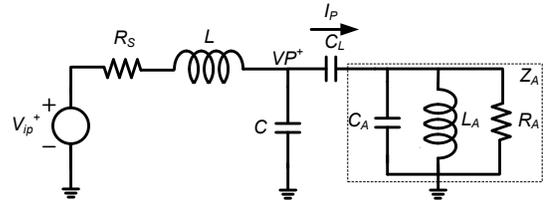,width=0.8\columnwidth}}
\caption{Circuit diagram of single-ended input low-power sub-GHz UWB transmitter.}
\label{uwbsingle}
\end{figure}

\hlchange{
The transmitter topology is based on the circuit principle depicted in Fig.~\ref{uwbsingle}.}
Unlike previously published works, the LPUT core contains a series LC network, comprising $R_S$, $L$ and $C$, which is driven by an impulse voltage source $Vip^+$.
The resistance $R_S$ is the series equivalent resistance of the LC network plus the output resistance of the voltage source.
Voltage $V\!P^+$ and current $ I_P$ are coupled into the antenna through capacitance $C_L$.
The antenna impedance, $ Z_A$, is modeled as a resistance, $ R_A$, in parallel with the antenna equivalent capacitance $ C_A$ and inductance $ L_A$, as shown in Fig.~\ref{uwbsingle}.
$ C_A$ and $ L_A$ are included in the antenna model since the antenna has a limited bandwidth, similar to a band-pass filter~\cite{Wouter2}.

Although the circuit from Fig.~\ref{uwbsingle} already generates an UWB pulse, the PSD of $V\!P^+$ still contains strong frequency components above $950 \unit{MHz}$ that couple into the antenna, violating the \hlchangeB{Federal Communications Commission (FCC)} regulation mask. To generate a pulse that complies with the FCC spectral mask, previous works perform power spectral density shaping by means of a filter~\cite{Wouter1, Wouter2}.
In addition, some designs rely on standard digital cell delays~\cite{Kin, Lei}. These methods introduce losses in the transfer function of the pulse shaping network in the transmitter or require high order filters to realize a steep roll-off near $950 \unit{MHz}$.

\begin{figure}
\centerline{\psfig{figure=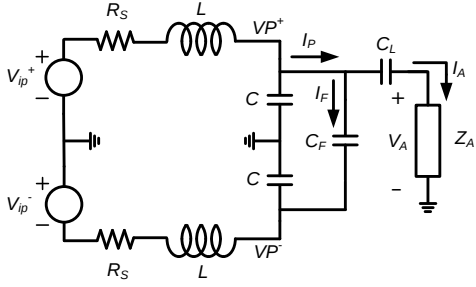,width=0.7\columnwidth}}
\caption{Circuit diagram of differential input low-power sub-GHz UWB transmitter.}
\label{uwbdiff}
\end{figure}

Fig.~\ref{uwbdiff} presents an input differential version of the circuit shown in Fig.~\ref{uwbsingle}.
The circuit has been duplicated and is driven by two impulse voltage sources, $Vip^+$ and $Vip^-$.
The voltages $V\!P^+$ and $V\!P^-$ have opposite signs as the network is driven pseudo-differentially.
The voltage and current signals at the antenna are single-ended since the currents $I_P$ and $I_F$ are subtracted in a single node.
Hence, the current at the antenna, $I_A$, is the difference between $I_P$ and $I_F$. $I_P$ is generated by $V\!P^+$ and $I_F$ is generated by the voltage difference $V\!P^+ - V\!P^-$ = $ V\!P_{DIF}$.

The impedance $Z_F = 1/(sC_F)$, ideally, is a short only for high frequency components.
A simple qualitative analysis can be made to understand the filtering effect of $Z_F$; 1: at low frequencies, $ Z_F$ is very high, therefore $ I_F = V\!P_{DIF}/Z_F$ is very small.
Hence, $I_A$ is not affected by $I_F$.
2: at high frequencies, $Z_F$ is very low.
Consequently, at high frequencies, the difference between $I_P$ and $I_F$, $I_A$, becomes very small.
This analysis leads us to understand that the upper limit of the PSD depends on $C_F$, the value of which can be selected to generate a pulse with a PSD that falls within the mask.

\begin{figure}
\centerline{\psfig{figure=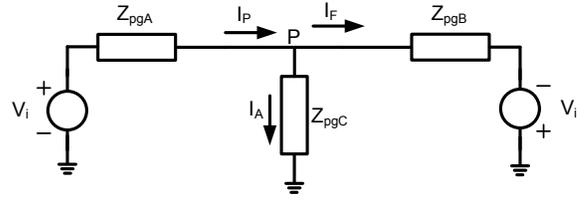,width=0.85\columnwidth}}
\caption{Circuit diagram of simplified LPUT as a T-network.}
\label{uwbcommnodePsimp}
\end{figure}

Fig.~\ref{uwbcommnodeP} shows the LPUT drawn as a quasi-symmetrical circuit, with quasi-symmetry seen from node P. Analyzing Fig.~\ref{uwbcommnodeP}, we can redraw the circuit as presented in Fig.~\ref{uwbcommnodePsimp}, which simplifies the analysis of currents $I_P$, $I_F$ and $I_A$ and the voltage at node P. Each network has an equivalent impedance, seen from node P to ground. The impedances can be described in the complex frequency domain as follows:

\begin{figure}
\centerline{\psfig{figure=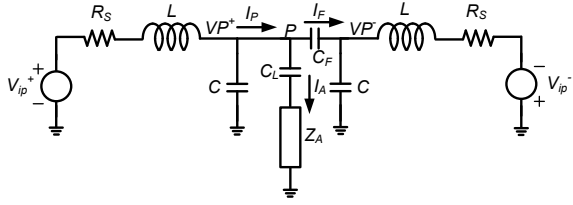,width=0.85\columnwidth}}
\caption{Circuit diagram of differential input LPUT as a quasi-symmetrical circuit.}
\label{uwbcommnodeP}
\end{figure}

\begin{equation}
 Z_{pgA} = \frac{(R_S + sL)} {(s^2LC + sR_SC + 1)},
\label{zpg}
\end{equation}

\begin{equation}
 Z_{pgB} = Z_{pgA} + Z_F,
\label{zpg1}
\end{equation}

\begin{equation}
 Z_{pgC} = Z_A + \frac{1} {(sC_L)},
\label{zpg2}
\end{equation}
and finally,
\begin{align}\label{VAtotal}
 V_A = &\left(\frac{ V_{ip}} { (s^2LC + sR_SC + 1)}\right) \\ \nonumber
&.\left(\frac{Z_F~Z_A} {Z_{pgC} Z_F + 2 Z_{pgA} Z_{pgC} + Z_{pgA} Z_F + Z_{pgA}^2}\right),
\end{align}
which is the voltage across the antenna as a function of the equivalent impedances and the input voltage.
From (\ref{VAtotal}), we can conclude that the antenna voltage is roughly zero if $ Z_F$ is very low. This analysis makes perfect sense as the network is fully symmetric and the voltage at node P becomes zero when the network is driven by a differential input voltage. Since $ Z_F$ is a capacitive reactance, the voltage at node P thus equals zero at high frequencies. In this design, $ C_F$ is chosen to set the voltage at node P to zero for frequencies above 1 GHz. 
\hlchange{The simulation results that validate the analysis are presented below.}

\begin{figure}
\centering
\includegraphics[width=0.9\columnwidth]{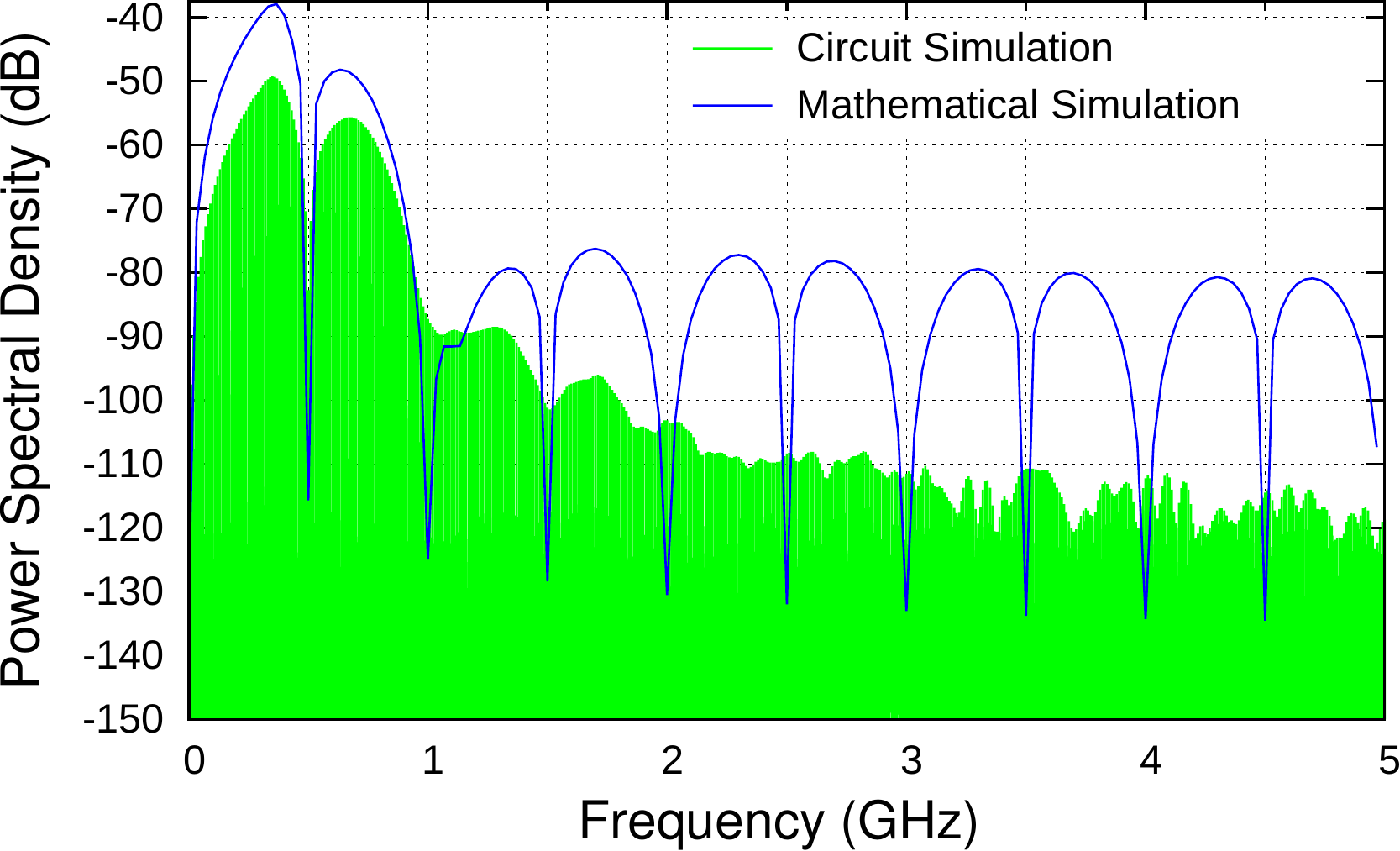}
\caption{PSD of circuit simulation and mathematical simulation.}
\label{PSDsim}
\end{figure}

Fig.~\ref{timesim} shows the time domain results of the circuit and mathematical simulation.
Fig.~\ref{PSDsim} presents the power spectral density of the simulated signals. 
The PSD of the signals from both simulations are very much alike and comply with the FCC mask. 
The difference between the two curves is that the mathematical simulation does not consider all losses that are present in the devices and in the PCB.
The circuit simulation already includes losses in the passive devices.
In addition, the circuit simulation includes the models of the micro-wave transistors that are used to implement the differential driver.
These transistors add more asymmetry since the gate-source voltages of the transistors are different when they drive the differential signal.

\begin{figure}
\centering
\includegraphics[width=0.9\columnwidth]{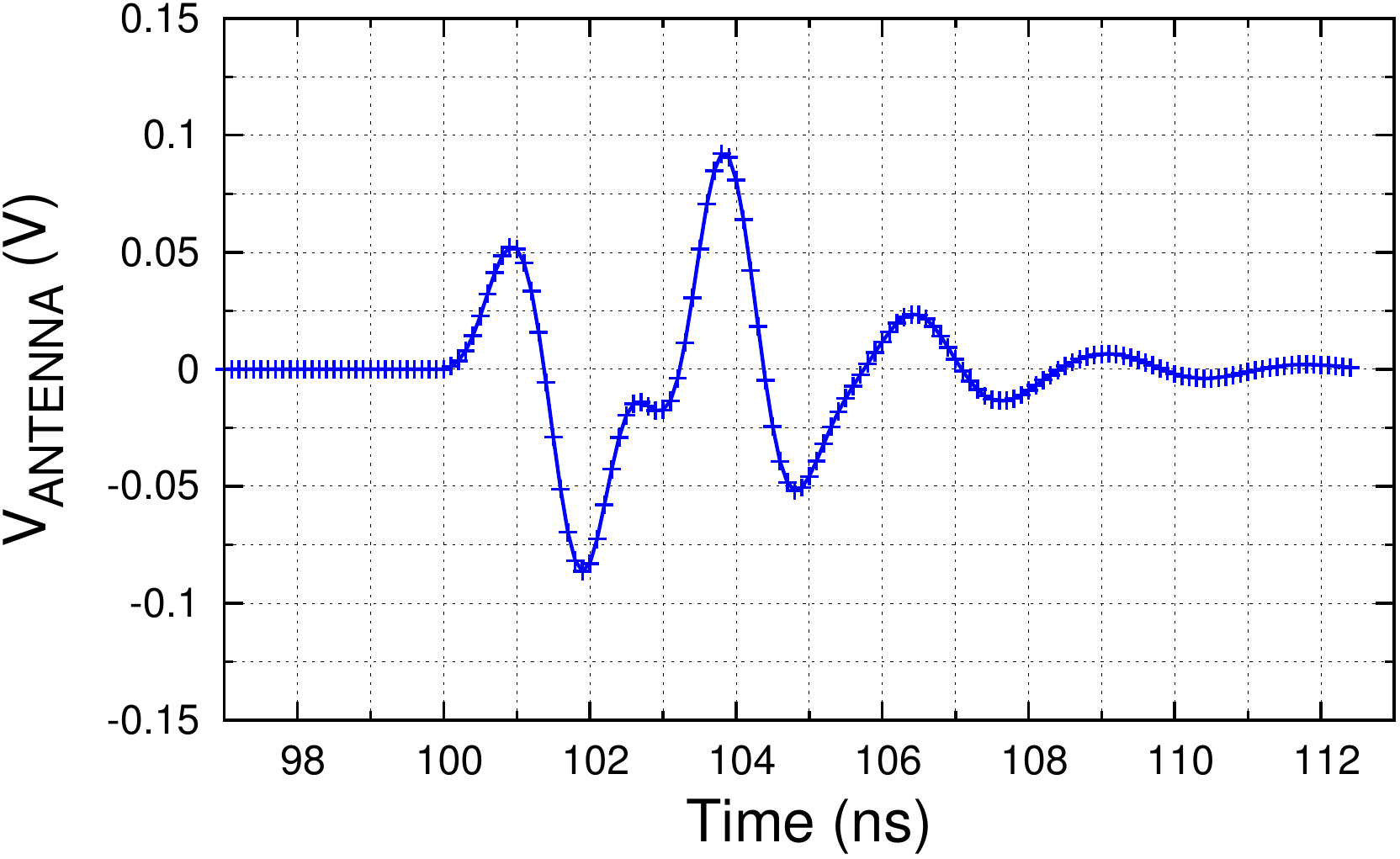}
\caption{Time domain result of circuit simulation and mathematical simulation.}
\label{timesim}
\end{figure}

\hlchange{
\subsection{Implementation and Results} \label{transmitter_example}
In this subsection, as a \hlchangeB{proof of principle}, an experimental implementation is described and measurement results are shown.}

\begin{figure}
\centerline{\psfig{figure=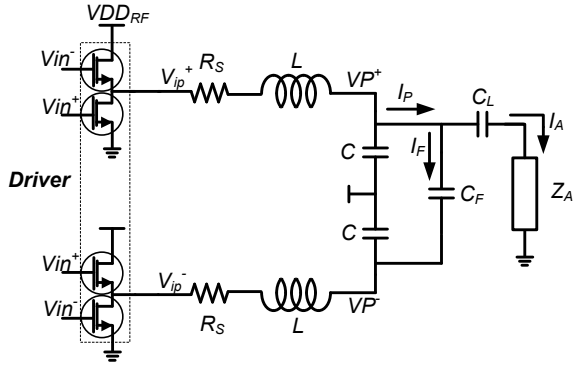,width=0.85\columnwidth}}
\caption{Circuit diagram of designed LPUT.}
\label{designedCD}
\end{figure}

The low-power sub-GHz UWB transmitter has been realized using high-speed discrete transistors, discrete (SMD) capacitors and on-PCB inductors in line with the circuit diagram shown in Fig.~\ref{designedCD}. The drivers are implemented using high-speed discrete transistors with low threshold voltages in a stacked topology that is suitable for low voltage operation ($100 \unit{mV}  \leq V\!D\!D_{RF}  \leq 150 \unit{mV}$) and offers high bandwidth.
The transistor that has been chosen for this design is the ATF551M4. The threshold voltage ($V_{TH}$) of this device is roughly $0.35 \unit{V}$. The input and output inductances and capacitances of this device are small enough to minimize dynamic power consumption and to keep the amount of high frequency spurs in the transmitted pulse small.

\begin{figure}
\centerline{\psfig{figure=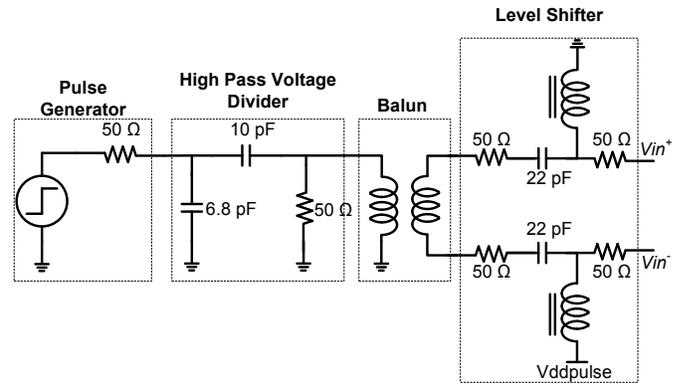,width=0.99\columnwidth}}
\caption{Circuit diagram of the stimulus generator.}
\label{stimulus}
\end{figure}

\begin{figure}
\centerline{\psfig{figure=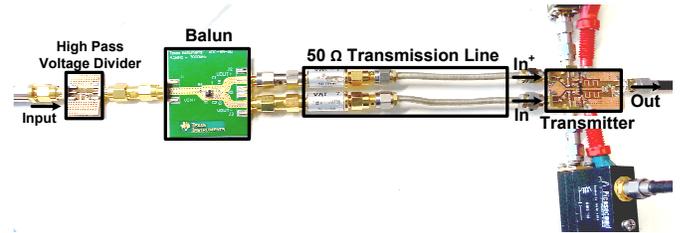,width=0.99\columnwidth}}
\caption{Test bench photograph of the LPUT.}
\label{testbench}
\end{figure}

\begin{figure}
\centerline{\psfig{figure=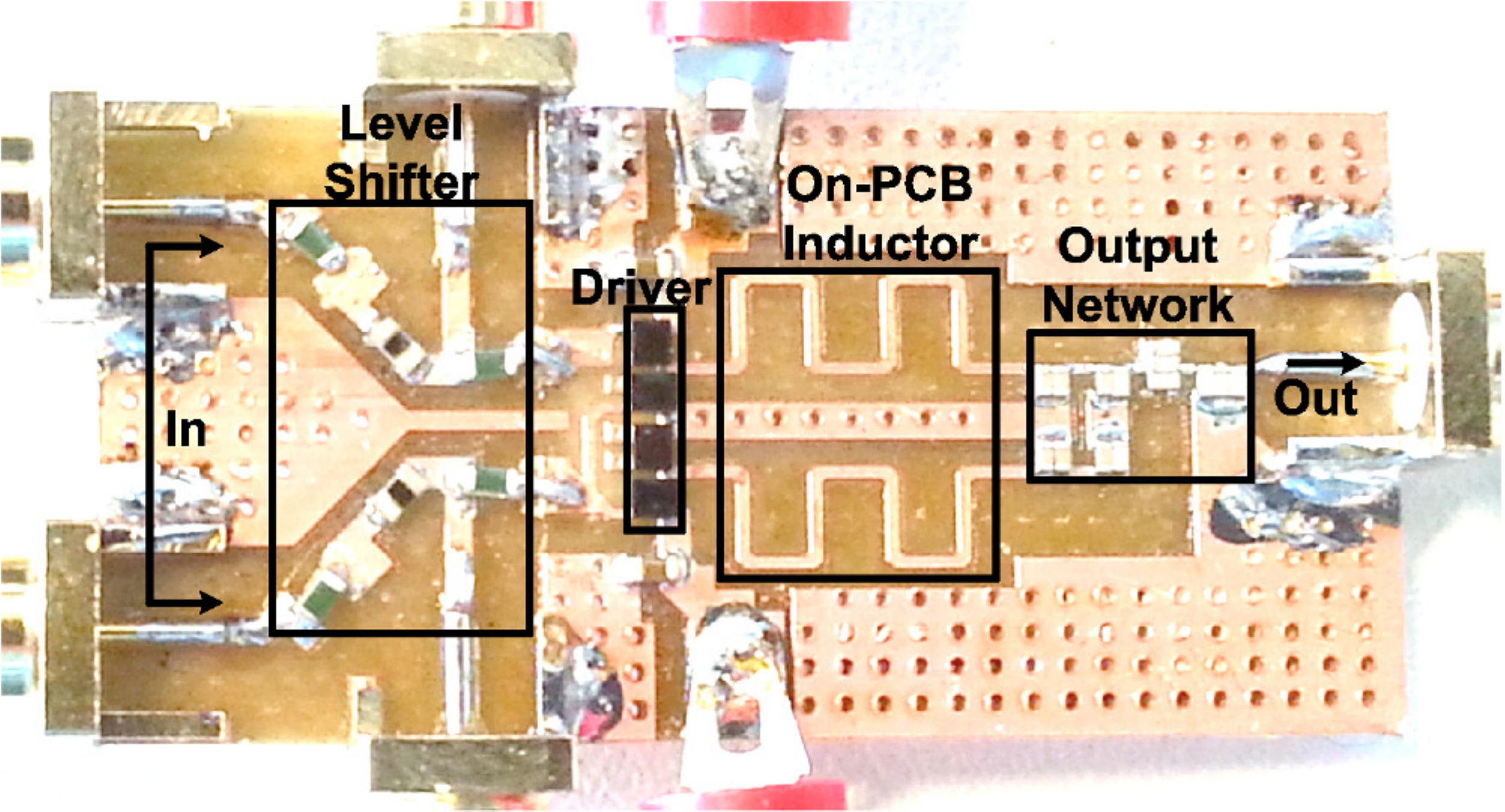,width=0.995\columnwidth}}
\caption{PCB photograph of the designed transmitter.}
\label{PCB_picture}
\end{figure}

For testing purposes only, a differential input stimulus generator has been implemented. The input stimulus generator comprises a pulse generator, a high pass voltage divider, a balun and a level shifter, as depicted in Fig.~\ref{stimulus}.
The level shifter is supplied by $V\!ddpulse$ that is $0.65 \unit{V}$ ($V_{TH}$ + $V\!D\!D_{RF}$) and thus large enough to drive both stacked transistors.
The photograph of the test bench is shown in Fig.~\ref{testbench}. Fig.~\ref{PCB_picture} shows a photograph of the PCB of the transmitter that includes the level shifter.
\hlchangeB{To measure the generated output signals, a spectrum analyzer and a oscilloscope were used.}

\begin{figure}
\centering
\includegraphics[width=\columnwidth]{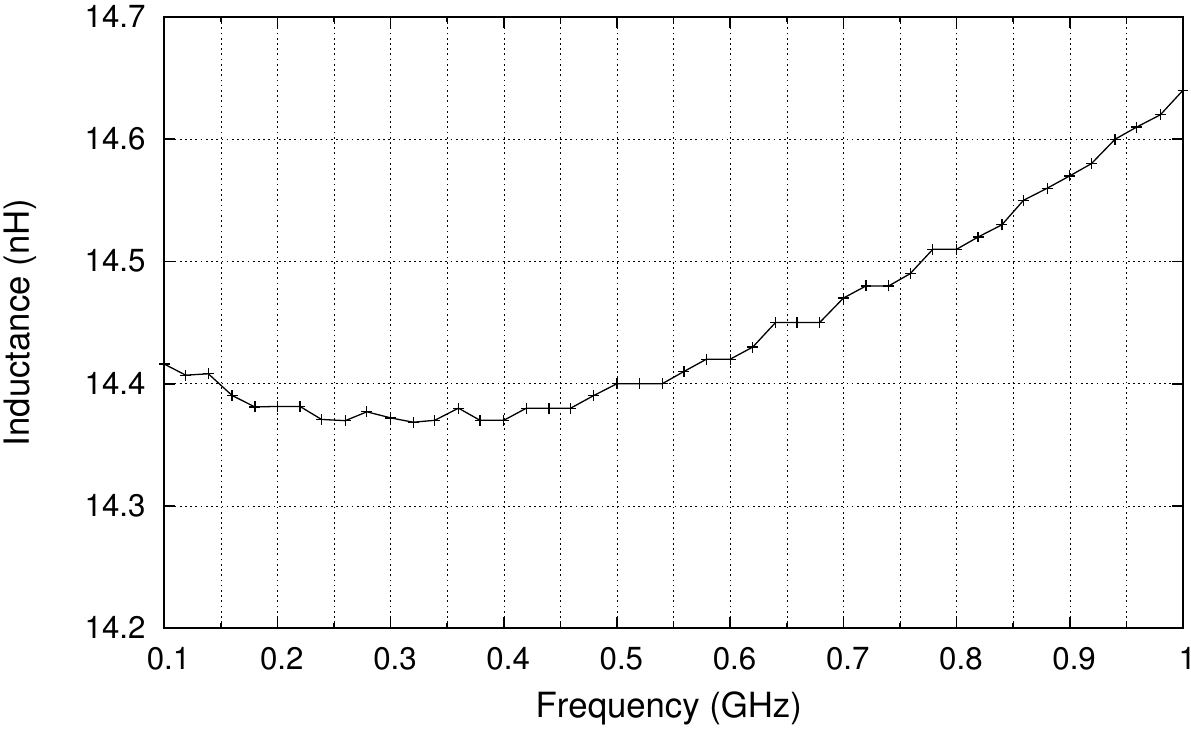}
\caption{Equivalent series resistance and inductance of the on-PCB inductor.}
\label{mom_inductor}
\end{figure}

The equivalent series resistance and inductance of the on-PCB inductor, extracted by means of electromagnetic simulation, are presented in Fig.~\ref{mom_inductor}.

\begin{figure} 
\centerline{\psfig{figure=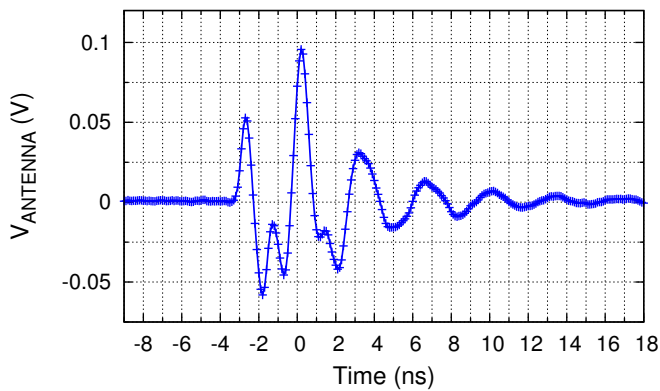,width=0.975\columnwidth}}
\caption{Measured output voltage waveform of the LPUT.}
\label{tranme}
\end{figure}

\begin{figure} 
\centerline{\psfig{figure=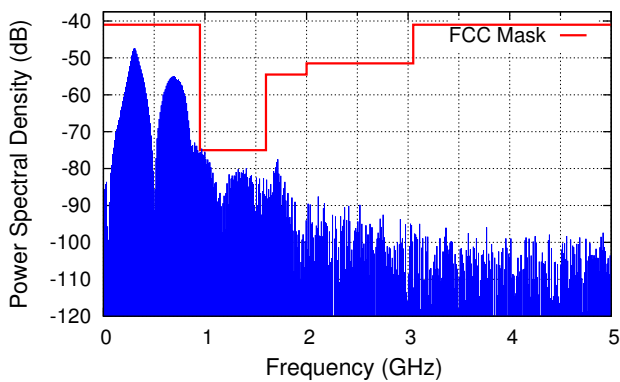,width=0.92\columnwidth}}
\caption{Measured output power spectral density of the LPUT.}
\label{PSDme}
\end{figure}

\begin{table*}
\begin{center}
\normalsize
\begin{threeparttable}
\renewcommand{\arraystretch}{1.325}
\caption{Performance Summary of the LPUT}
\label{TABH}
\begin{tabular}{c||c|c||c}
\hline
\textbf{Specifications} &~\textbf{Value} &~\textbf{Specifications} &~\textbf{Value}\\
\hline

\hline
Frequency Band (GHz) & 0.25-0.75 & $\rm V_{peak-peak}$ (V) & 0.14 \\
\hline
Power (mW) & 0.280 & $\rm V_{peak-peak}/Supply~Voltage$ ($\%$) & 93.3\\
\hline
Energy/Pulse (pJ/pulse) & 85 & PRF (MHz) & 3.3\\
\hline
Supply Voltage (V) & 0.15/0.65$^\dag$  & Roll-off (dB/octave) &  25\\
\hline
\end{tabular}
\item [$\dag$] Voltage supply of the level shifter, for testing purposes.
\end{threeparttable}
\end{center}
\end{table*}

Fig.~\ref{tranme} and Fig.~\ref{PSDme} show the measured output voltage waveform of the LPUT and the corresponding power spectral density for $V\!D\!D_{RF} = 0.15 \unit{V}$, respectively.
In Fig.~\ref{PSDme}, some high frequency components ($> 1 \unit{GHz}$) can be observed while in Fig.~\ref{PSDsim} high frequency components are attenuated.
\hlchangeB{Due to an unbalance between the differential inductors and capacitors (L and C in Fig.~\ref{uwbcommnodeP}), an unbalance in the transient response can be observed. The output voltage reaches $-0.06 \unit{V}$, while it reaches $-0.1 \unit{V}$ in the simulations.}

The main difference between the measured and the mathematical results is the symmetry of the circuit. The circuit developed on PCB is not ideally symmetrical, therefore the roll-off is less steep than that of the mathematical model. The PSD of the transmitted signal presents a 25 dB decay between 500 MHz and 1 GHz.
\textcolor{black}{
This steep decay is a consequence of the differential to single ended conversion of the circuit.
It comes from the fact that the neutral point of the differential circuit behaves as a ground for very high frequencies.
The steep decay in the PSD is an important characteristic of the circuit that allows to keep the transmitted signal within the FCC mask \cite{FCC_reference}.
}

\begin{figure} 
\centering
\subfigure[]{\psfig{figure=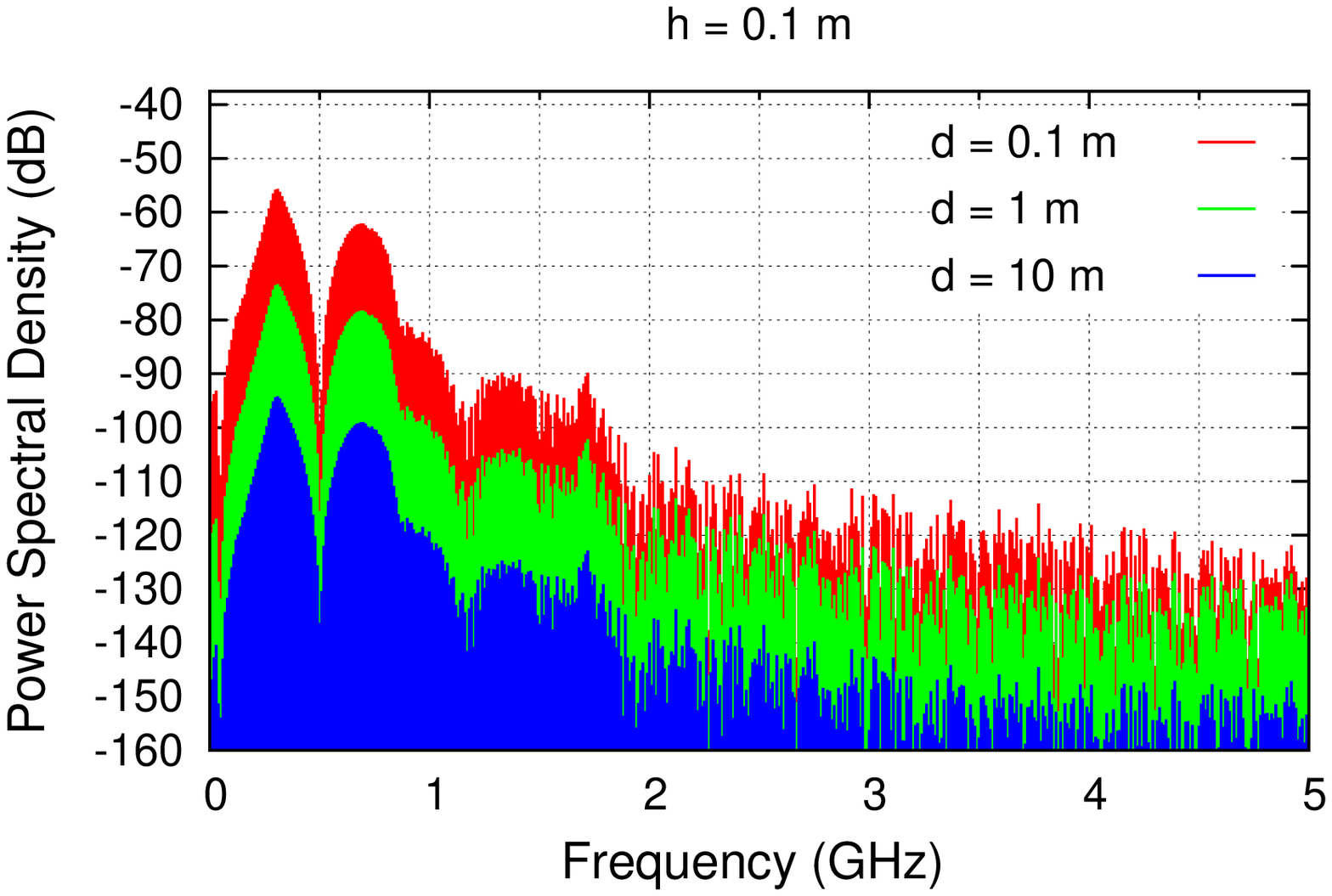,width=0.82\columnwidth}
\label{PSD01m}
}
\subfigure[]{\psfig{figure=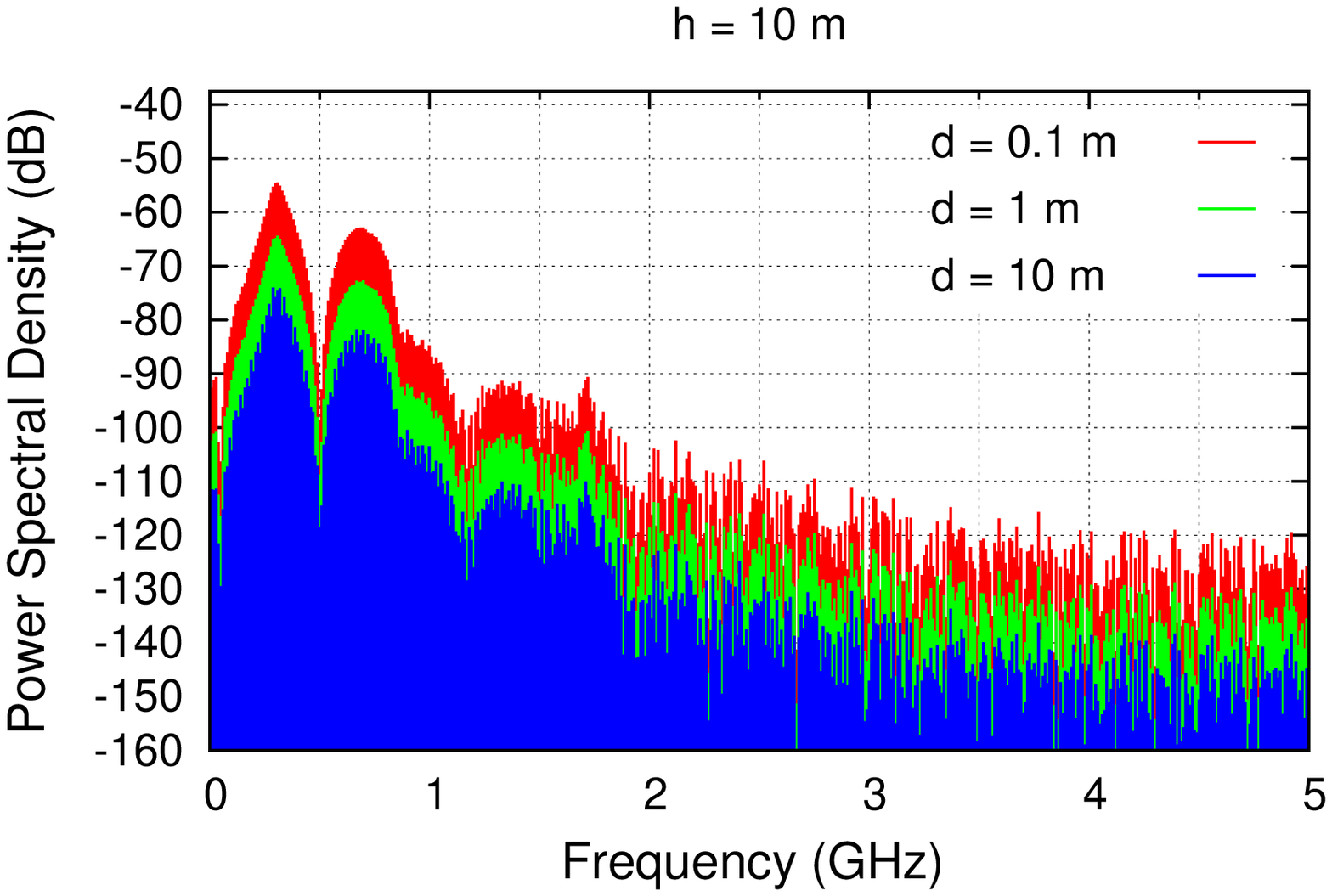,width=0.82\columnwidth}
\label{PSD10m}
}
\caption{Estimated PSD of the LPUT including path loss and ground reflection for (a) $h = 0.1 \unit{m}$ and (b) $h = 10 \unit{m}$.}
\label{fig:psdlput}
\end{figure}

Fig. \ref{fig:psdlput} presents the estimated PSD of the transmitted impulse (of Fig.~\ref{PSDme}) after taking into account path loss, ground reflection and antenna/receiver matching.
The path loss and ground reflection are modeled according to \cite{Promwong, Friis} and simulated using the measured transmitted impulse as input.
The transmitter-receiver separation distances ($d$) considered in the analysis are $0.1 \unit{m}$, $1 \unit{m}$, and $10 \unit{m}$ with the transmitter and receiver in the same height ($h$) of $0.1 \unit{m}$ and $10 \unit{m}$; the ground reflection coefficient is $-1$. 
Fig. \ref{fig:psdlput} also shows that the PSD peak power decreases with $d$ and increases with $h$.
This behavior can be explained by the fact that at lower height ($h = 0.1 \unit{m}$) the ground reflected signals are stronger at the receiving antenna with $180^\circ$ phase.
On the other hand, if the antenna is higher ($10 \unit{m}$) the reflected signal is attenuated and its effects are minimized.

\hlchange{
\subsection{Discussion} \label{transmitter_discussion}}
The performance summary of the LPUT is presented in Table \ref{TABH}.
The power consumption of the UWB transmitter is only $0.28$\unit{mW}. This value is the addition of both the driver and the level shifter power consumption, and it is much lower lower than the typical power consumption of the state-of-the-art transmitters presented in Table \ref{tab:txpwr}.
The transmitter, usually, is the most power hungry block in a WSN. Therefore, with the proposed UWB transmitter, the overall power consumption of the WSN is drastically reduced. This allows to bring the two curves of Fig. \ref{power_scenario} closer to each other.
\hlchange{The power consumption of the transmitter can be further reduced by employing transistors with a lower input capacitance and lower threshold voltage, i.e., using a more advanced technology node.
It is worth mentioning that the transmitter presented in this section requires a wide-band antenna to operate, while the receiver and the RF-DC converter presented previously require a narrow-band antenna. This means that it is not optimal for the UWB transmitter and the other circuits to share the same antenna.
Alternatively, backscattering communication can be used to transmit and receive data through the same antenna. 
If the transmitter has its own antenna (Scenarios c) and e)), a higher appropriate frequency can be used for transmission, resulting in higher data rate but at the expense of higher power consumption.
}

\section{Summary and Conclusions}
\label{sec:conc}
The circuit techniques presented in this paper allow to have both energy and bidirectional data transfer to a sensor node in a low-power and energy-efficient manner. 
\hlchange{Five different scenarios of antenna configurations in which a WSN can operate and how those circuit techniques can be applied in those scenarios have been described. 
The three fundamental blocks considered are: RF to DC converter, data receiver and data transmitter.
With repect to the RF-DC conversion, a voltage boosting network combined with a 5-stage on-chip rectifier and its measurement results were presented.
A DC-DC converter was presented as an alternative to a multistage on-chip rectifier.
The implemented buck-boost DC-DC converter employs an MPPT technique that estimates the input power, and adjusts the equivalent input resistance of the DC-DC converter in order to maximize the power extracted from the rectifier, allowing for efficient harvesting across a larger range of available input power.
Regarding both the RF-DC conversion and the data receiver, we showed that co-design of the antenna and electronics leads to better performance for the overall system.
For the data receiver, the co-design technique leads to better performance in terms of NF and power efficiency, and for the RF-DC converter it leads to better power conversion efficiency by boosting the rectifier input voltage.
In order to achieve low-power data transmission, the task that usually consumes the most power in a WSN, a novel low-power sub-GHz UWB transmitter was presented along with its measurement results.
Throughout the paper a quantified comparison with some relevant prior art, has been carried out, proving that the power gap present in the state of the art WSNs can be drastically reduced, or even disappear.}

\bibliographystyle{IEEEtran}
\bibliography{IEEEabrv,refs}

\vspace*{-2\baselineskip}
\begin{IEEEbiography}
    [{\includegraphics[width=1in,height=1.25in,clip,keepaspectratio]{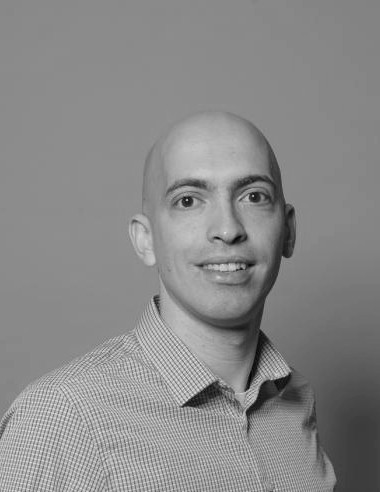}}]{Gustavo C. Martins}
(S'11) was born in Jacupiranga, Brazil, in 1988. He received the B.Eng. degree from
Universidade de S\~ao Paulo, S\~ao Carlos, Brazil, in 2010
and the M.Sc. degree from Universidade Federal de Santa Catarina, Florian\'opolis, Brazil, in 2013.
He is currently a Ph.D. candidate at the Section Bioelectronics of Delft University of Technology, Delft, The Netherlands.
His research interests include wireless power transfer and low-power analog IC design.
\end{IEEEbiography}

\begin{IEEEbiography}
    [{\includegraphics[width=1in,height=1.25in,clip,keepaspectratio]{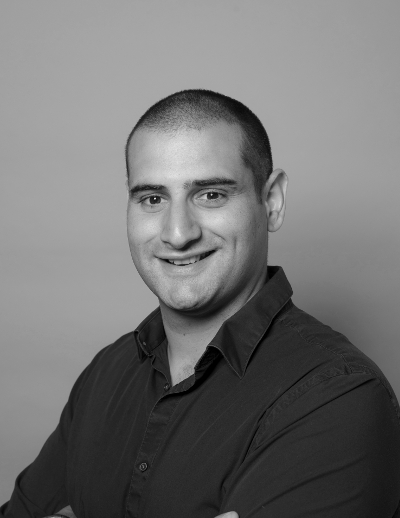}}]{Alessandro Urso}(S'16) 
was born in the province of Lecce, Italy, in 1991. He received the Bachelors degree and the M.Sc. degree (cum Laude) in Electronic and Telecommunications engineering from the University of Ferrara, Italy in 2013 and 2015, respectively. He is currently a Ph.D. candidate at the Section Bioelectronics of Delft University of Technology, Delft, The Netherlands. His research interests include the design of power efficient neural stimulator  as well as the design of switched capacitor DC-DC converter for energy harvesting application.
\end{IEEEbiography}

\begin{IEEEbiography}
    [{\includegraphics[width=1in,height=1.25in,clip,keepaspectratio]{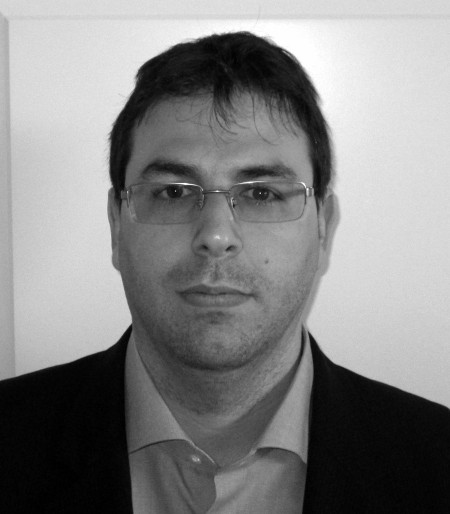}}]{Andr\'e L. Mansano}
(S'11) was born in Jaboticabal,
Brazil, in 1982. He received the B.Eng. degree from
Universidade Estadual Paulista, S\~ao Paulo, Brazil,
and the M.Sc. degree from Universidade Estadual
de Campinas, S\~ao Paulo, Brazil, in 2006 and 2009,
respectively. In 2016 he received his PhD degree from Delft University of Technology.
In 2007, he joined Freescale Semiconductor as a
Analog Mixed-Signal Designer. He designed analog
IPs for automotive and general purpose 8 and 32-bit
microcontrollers. In addition to the designed IPs for commercial products, 
he has authored two patents and four papers. In November 2010, he joined Delft University of Technology, Delft,
The Netherlands, as a Guest Researcher. Currently he works at Philips Research in Eindhoven as analog mixed signal IC designer.
His research interests include wireless energy harvesting, wirelessly powered sensors, very-low-power and low-noise amplifiers
and low-power wireless communication.
\end{IEEEbiography}

\begin{IEEEbiography}
    [{\includegraphics[width=1in,height=1.25in,clip,keepaspectratio]{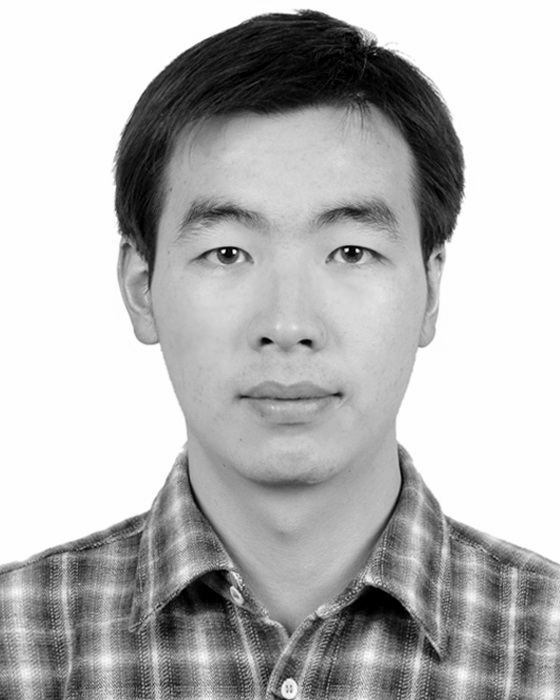}}]{Yao Liu}
was born in Hubei, China, in 1987. He received the M.Sc. and B.Sc. degrees in Electrical Engineering from Huazhong University of Science and Technology, Wuhan, China, in 2008 and 2011, respectively. He is currently working towards the Ph.D. degree in the Section Bioelectronics of the Department of Microelectronics at the Faculty of Electrical Engineering, Mathematics, and Computer Science, Delft University of Technology. His research interests include low power wireless communication circuits design for biomedical applications.
\end{IEEEbiography}

\begin{IEEEbiography}
    [{\includegraphics[width=1in,height=1.25in,clip,keepaspectratio]{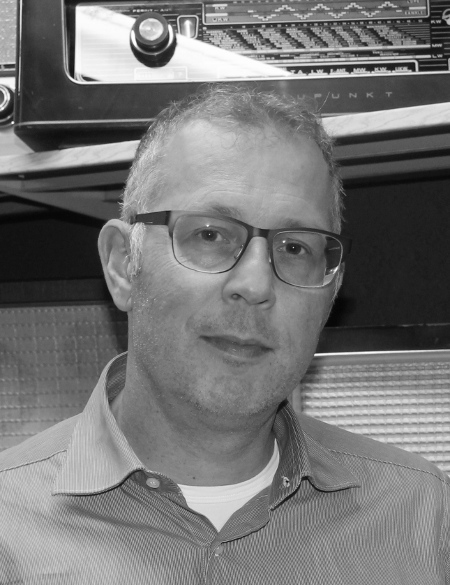}}]{Wouter A. Serdijn}
(M'98, SM'08, F'11) was born in Zoetermeer ('Sweet Lake City'), the Netherlands, in 1966. He received the M.Sc. (cum laude) and Ph.D. degrees from Delft University of Technology, Delft, The Netherlands, in 1989 and 1994, respectively. Currently, he is a full professor in bioelectronics at Delft University of Technology, where he heads the Section Bioelectronics, and a visiting honorary professor at University College London, in the Analog and Biomedical Electronics group.

His research interests include integrated biomedical circuits and systems for biosignal conditioning and detection, neuroprosthetics, transcutaneous wireless communication, power management and energy harvesting as applied in, e.g., hearing instruments, cardiac pacemakers, cochlear implants, neurostimulators, portable, wearable, implantable and injectable medical devices and electroceuticals.
 
He is co-editor and co-author of 10 books, 8 book chapters, 2 patents and more than 300 scientific publications and presentations. He teaches Circuit Theory, Analog Integrated Circuit Design, Analog CMOS Filter Design, Active Implantable Biomedical Microsystems and Bioelectronics. He received the Electrical Engineering Best Teacher Award in 2001, in 2004 and in 2015.
 
He has served, a.o., as General Co-Chair for IEEE ISCAS 2015 and for IEEE BioCAS 2013, Technical Program Chair for IEEE BioCAS 2010 and for IEEE ISCAS 2010, 2012 and 2014, as a member of the Board of Governors (BoG) of the IEEE Circuits and Systems Society (2006-2011), as chair of the Analog Signal Processing Technical Committee of the IEEE Circuits and Systems society, and as Editor-in-Chief for IEEE Transactions on Circuits and Systems-I: Regular Papers (2010-2011). Currently, he is a member of the Steering Committee and an Associate Editor of the IEEE Transactions on Biomedical Circuits and Systems (T-BioCAS)
 
Wouter A. Serdijn is an IEEE Fellow, an IEEE Distinguished Lecturer and a mentor of the IEEE. In 2016, he received the IEEE Circuits and Systems Society Meritorious Service Award.
\end{IEEEbiography}

\vfill

\vfill

\end{document}